\documentclass[12pt]{article}

\usepackage{titlesec}
\usepackage[plain]{algorithm}
\usepackage{algpseudocode}
\usepackage{amsfonts}
\usepackage{amsmath}
\usepackage{amssymb}
\usepackage{amsthm}
\usepackage[toc,page]{appendix}
\usepackage[american]{babel}
\usepackage{bm}
\usepackage[format=hang,justification=raggedright,margin=35pt,font=bf,singlelinecheck = false]{caption}
\usepackage{color}   %May be necessary if you want to color links
\usepackage{courier}
\usepackage[useregional=numeric]{datetime2}
\usepackage{etoolbox}
\usepackage{environ}
\usepackage{extramarks}
\usepackage{geometry}
\usepackage{fancyhdr}
\usepackage{lipsum}     % for sample text
\usepackage{mdframed}   % for framing
\usepackage[neverdecrease]{paralist}
\usepackage{textcomp}
\usepackage[nottoc]{tocbibind}
\usepackage[breakable, theorems, skins]{tcolorbox}
\usepackage{tikz}
\usepackage[normalem]{ulem}
\usepackage{xcolor}     % for colour
\usepackage{pdflscape}
\usepackage{booktabs}
\usepackage{caption}
\usepackage{longtable}[=v4.13]
\usepackage{graphicx} 
\usepackage{array}
\usepackage{multirow}
\usepackage{wrapfig}
\usepackage{float}
\usepackage{colortbl}
\usepackage{pdflscape}
\usepackage{tabu}
\usepackage{threeparttable}
\usepackage{threeparttablex}
\usepackage{makecell}
\usepackage{setspace}
\usepackage{float}
\usepackage{supertabular}
\usepackage{IEEEtrantools}
\usepackage{subcaption}
\usepackage{tocloft,calc}
\usepackage[pdfauthor={Jonathan Mendelson and Michael R. Elliott}, pdftitle={Bayesian optimal sample design for surveys with heteroscedasticity}, pdfkeywords={sample allocation, optimal allocation, stratified sampling, Bayesian decision theory, measure of heteroscedasticity}]{hyperref}

\usepackage{csquotes}

\usepackage[authordate, backend=biber, natbib=true, giveninits=true]{biblatex-chicago}

\hypersetup{
	colorlinks=true, %set true if you want colored links
	linktoc=all,     %set to all if you want both sections and subsections linked
	linkcolor=blue,  %choose some color if you want links to stand out
	citecolor=blue,
	urlcolor=blue
}

\renewcommand{\part}[1]{\textbf{\large Part \Alph{partCounter}}\stepcounter{partCounter}\\}

\DeclareFieldFormat{doi}{%
	DOI\addcolon\space
	\ifhyperref
	{\href{https://doi.org/#1}{\nolinkurl{https://doi.org/#1}}}
	{\nolinkurl{#1}}%
}

\usetikzlibrary{shapes,arrows}
\tikzstyle{decision} = [diamond, draw, fill=blue!20, 
text width=4.5em, text badly centered, node distance=3cm, inner sep=0pt]
\tikzstyle{block} = [rectangle, draw, fill=blue!20, 
text width=5em, text centered, rounded corners, minimum height=4em]
\tikzstyle{line} = [draw, -latex']
\tikzstyle{cloud} = [draw, ellipse,fill=red!20, node distance=3cm,
\textbf{}   minimum height=2em]

\newcommand{\tab}[1]{
	\hspace{.2\textwidth}\rlap{#1}}
\useunder{\uline}{\ul}{}

\newcommand{\appendixpagenumbering}{
	\break
	\pagenumbering{arabic}
	\renewcommand{\thepage}{\thesection.\arabic{page}}
}

\title{Bayesian optimal sample design for surveys with heteroscedasticity}
\author{Jonathan Mendelson\thanks{Jonathan Mendelson is Research Statistician at the US Bureau of Labor Statistics, Suitland, MD 20746, USA.} \and Michael R. Elliott
	\thanks{Michael R. Elliott is Professor of Biostatistics and Research Professor of Survey Methodology at the University of Michigan, Ann Arbor, MI 48109, USA.\\ \\
		The authors would like to thank Joseph Sedransk, Eric Slud, Richard Valliant, Partha Lahiri, and Roderick Little, whose suggestions greatly improved the quality of this manuscript. Any opinions expressed in this paper are those of the authors and do not constitute policy of the Bureau of Labor Statistics. \\ \\
		Address correspondence to Jonathan Mendelson, Office of Survey Methods Research, U.S. Bureau of Labor Statistics, 4600 Silver Hill Road, Suitland, MD 20746, USA;
		E-mail: \href{mailto:mendelson.jonathan@bls.gov}{mendelson.jonathan@bls.gov}. }}
\date{July 2026}

\renewcommand{\appendixpagename}{Bayesian optimal sample design for surveys with heteroscedasticity: appendix}

\begin{document}
	\maketitle
	\linespread{1}
	\selectfont

\begin{abstract}

        We develop a Bayesian optimal sample allocation approach for stratified sampling in heteroscedastic populations.
        Existing optimal allocation theory typically assumes knowledge of certain design parameters (e.g., strata variances) that may be unknown, leading practitioners to substitute in survey-based estimates when planning samples, often without considering the effects of this substitution on sample efficiency.
        Bayesian decision theory for optimal experimental design avoids such substitutions and can be applied to sample allocation.
        Bayesian sample optimization methods were studied heavily from the mid-1960s through early 1980s, but have been overlooked since, in spite of modern computing advances that have facilitated a proliferation of Bayesian methods in other areas of statistics.
        A limitation of this early Bayesian sample design work is that it did not accommodate heteroscedastic error structures, which underlie commonly used ratio estimation models.
        Our paper, which optimizes the design under a univariate regression model with heteroscedastic errors, addresses this limitation of earlier work, while illustrating the Bayesian approach to design.
        We identify the optimal Bayesian allocation under our model, then compare performance of the proposed Bayesian sampling strategy with that of key design-based and model-assisted alternatives across several settings, finding that the proposed methods do as well or better than the alternatives under the scenarios considered.
        We apply our methods in analyzing revenues of public charities, using publicly available IRS Form 990 data.
        
\end{abstract}


\pagebreak

\section{Introduction}
\label{sec:ch2:intro}

Survey statisticians have long been interested in optimal designs, wherein optimality refers to minimizing some objective function (e.g., variance of a finite population total) subject to certain constraints (e.g., maximum costs or sample size). Optimizing survey designs has a long history dating back
at least to \citet{Neyman1934}, who recognized that sampling
strata proportional to their standard deviation was more
efficient than sampling proportional to their population
sample size if strata variances differ.
Neyman allocation was subsequently extended to account
for differential data collection costs across strata \citep[e.g.,][]{Cochran1977}.

Typically, the theoretically optimal sample allocations are conditional on population characteristics that are unknown (e.g., strata variances for Neyman allocation).
In practice, such design parameters are often estimated from prior surveys, after which the estimates are substituted in place of the true quantities in determining the design.

Despite vast spending on sample surveys \citep{OMB2017}, there is only a modest amount of recent literature on designing samples in the presence of uncertain design parameters.
From a frequentist perspective,  \citet{Clark2013} proposed a statistical learning approach for allocating stratified random samples, which led to improvements, although Clark noted that Bayesian methods may be more efficient.
There is a sizable Bayesian literature on optimal survey sample design, largely published from the mid-1960s through early 1980s.
A common approach entails deriving the posterior loss (e.g., posterior variance) conditional on the data, averaging over the predictive distribution for the data, and then selecting the sample design that minimizes this expected posterior loss \citep[e.g.,][]{Draper1968a,Rao1972}.
Given that the Bayesian optimal sample design literature predates recent modern computing advances, this approach typically requires a closed form for the expected posterior loss, which might not be possible for some models.
Further, the models employed may sometimes be unrealistic.
For example, \citet{Draper1968a} consider optimal allocation for stratified random sampling, but employ a normal likelihood with fixed strata means and variances, which does not accommodate heteroscedastic errors.
In contrast, \citet{Rao1972} examine optimal design for categorical data via a Dirichlet-multinomial model, which may be useful in several contexts, but not necessarily for settings in which distributions are continuous specification and possibly skewed.

We examine stratified random sampling for surveys wherein the stratum means and possibly variances are functions of a known population-level auxiliary variable.
We assume that some pilot data set is available for estimating parameters needed for designing a main study.
Our aim to identify optimal sample allocations for given strata; this is in contrast to a focus on identifying optimal strata boundaries for a size-related measure \citep[e.g., ][]{Wright1983,Dalenius1959,Rivest2002}.
Although our focus is on establishment surveys that often have large degrees of heterogenity \citep{Henry2009,Smith2003}, the methods developed here are relevant to any settings with stratified sampling with heterogeneous variances within strata.

We formulate our problem using Bayesian decision theory on optimal experimental design \citep[e.g.,][]{Lindley1972}, which is a general and flexible approach that can accommodate analyses such as those of \citet{Draper1968a} and \citet{Rao1972}, and which accounts for uncertain estimates of design parameters.
We assume that inferences are made using a Bayesian framework for finite population inference, following \citet{Ericson1969}.
We derive approximate analytical results for the expected posterior variance for the finite population mean, then identify an allocation using standard convex optimization methods.

The remainder of this manuscript is organized as follows.
Background is provided in Section \ref{sec:ch2:Background}, and includes attention to stratified sample allocation under our assumed model, sampling with uncertain design parameters, and Bayesian optimal experiment design.
Section \ref{sec:ch2:BayesSTSRS_analytic_results} presents analytical results and shows how to identify the Bayesian optimal design for estimating the finite population mean, assuming the proposed model and a squared error loss function.
Section \ref{sec:ch2:Simulation_design} outlines simulation methods for comparing the proposed allocations with alternatives across 90 artificial populations.
Section \ref{sec:ch2:sim_results} presents simulation findings.
Section \ref{sec:ch2:NCCS_simulation} repeats the simulation in an application to the log-revenues of public charities in the United States.
Section \ref{sec:ch2:discussion} summarizes results, discusses limitations, and suggests future directions.

\section{Background}
\label{sec:ch2:Background}

\subsection{Design-based approaches for optimal allocation using ratio estimators}
\label{subsec:background:design_based_sampling}

\subsubsection{Optimal allocation for the separate ratio estimator}
\label{subsec:Cochran_allocation_separate_ratio_estimator}
We consider estimation of a population total $Y=\sum_{h=1}^H\sum_{i=1}^{N_h}Y_{hi}$, in a context where we know values for a related variable, $\{X_{hi}\}$, for the population.  Our sample is chosen via an  stratified simple random sampling (STSRS) design.

Assume that within each stratum, the $\{Y_{hi}\}$ and $\{X_{hi}\}$ are correlated, but with correlations that may vary by strata.
For convenience, let $\{y_{hi},x_{hi}\}$ refer to the values of the $n=\sum_{h=1}^{H}n_h$ sampled units for variables $\{Y_{hi},X_{hi}\}$.
Let $\bar{y}_h=\frac{1}{n_h}\sum_{i=1}^{n_h}y_{hi}$ and $\bar{x}_h=\frac{1}{n_h}\sum_{i=1}^{n_h}x_{hi}$ denote the stratum $h$ sample means and $Y_{h}=\sum_{i=1}^{N_h} Y_{hi}$ and $X_{h}=\sum_{i=1}^{N_h} X_{hi}$ denote the stratum population totals of our outcome and auxiliary variables, respectively.
Let $R_h=\frac{X_h}{Y_h}$ be the ratio of stratum population totals.
This scenario commonly employs the separate ratio estimator of the population total, 
\begin{align}
	\hat{Y}_{Rs} &= \sum_{h=1}^{H}\hat{R}_h X_h 
	\\ &= \sum_{h=1}^{H}\frac{\bar{y}_h}{\bar{x}_h}X_h, \label{eqn_SR_estimator}
\end{align}
\noindent termed \textit{separate} (versus \textit{combined}) by virtue of allowing the strata slopes to vary.
\citet[][sec. 6.14]{Cochran1977} indicated that the optimal allocation for the separate ratio estimator is 
\begin{align}
	n_h \propto \frac{N_h S_{dh}}{\sqrt{c_h}} \label{eqn:Cochran_ratio_allocation}
\end{align}
\noindent where $S^2_{dh}=\frac{1}{N_h-1}\sum_{i=1}^{N_h}d^2_{hi}$ and $d_{hi}=Y_{hi}-\frac{Y_h}{X_h} X_{hi}$ is the deviation of $Y_{hi}$ from $\frac{Y_h}{X_h}X_{hi}$.
This allocation is derived by selecting $\{n_h\}$ to minimize the approximate variance of the separate ratio estimator, $\mathrm{Var}\left(\hat{Y}_{Rs}\right)\overset{.}{=}\sum_{h=1}^{H}N_h^2(1-f_h)\frac{S_{dh}^2}{n_h}$, the latter of which assumes large sample sizes in each stratum.

\citeauthor{Cochran1977} did not explicitly provide an estimator for $S_{dh}$, which he suggested is difficult to speculate about in the context of sample planning (p. 172).
Instead, he used the theoretically optimal allocation of Equation (\ref{eqn:Cochran_ratio_allocation}) to justify approximate solutions tailored to the type of population, based on the relationship between $S_{dh}$ and $\bar{X}_h$.
For scenarios in which the ratio estimate is a best linear unbiased estimate (BLUE), $S_{dh}$ will be roughly proportional to $\sqrt{\bar{X}_h}$, suggesting an allocation of $n_h \propto N_h \sqrt{\bar{X}_h}/\sqrt{c_h}$.
In populations wherein the $d_{hi}$ are more closely proportional to $\bar{X}_h^2$, Cochran suggested an allocation of $n_h \propto N_h \bar{X}_h/\sqrt{c_h}$.

Although not mentioned by Cochran, if prior data are available, an alternative to these rules of thumb would be to employ a plug-in allocation of the form $n_h \propto N_h \hat{S}_{dh}/\sqrt{c_h}$, where $\hat{S}_{dh}$ is a sample-based estimate from some previous data.
One such estimator is 
\begin{align}
	\hat{S}^2_{dh} = \frac{1}{m_h-1}\sum_{i\in s_{1h}}\left(y_{hi}-\hat{R}_h x_{hi}\right)^2, \label{eqn:Cochran_plugin_SS_estimator}
\end{align} where $\hat{R}_h=\frac{\sum_{i\in s_{1h}}y_{hi}}{\sum_{i\in s_{1h}} x_{hi}}$ is a ratio estimate from a pilot sample in stratum $h$ of size $m_h$, and where $s_{1h}$ refers to the pilot sample in stratum $h$.
Although possibly reasonable in some contexts, this estimate of $S^2_{dh}$ has bias of order $1/m_h$ \citep[][32]{Cochran1977}, suggesting worsened performance when using smaller pilot samples.

\subsubsection{Heteroscedasticity and optimal sample design}
\label{subsec:background:heteroscedasticity}
Although our ultimate focus is on Bayesian optimal design, here, we review frequentist methods for handling heteroscedasticity in sample design.
Our eventual focus will be on optimal sample allocation for STSRS designs, assuming a stratified population model that follows $Y_{hi}|(X_{hi},\alpha_h,v_h,b) \sim N\left(\alpha_h X_{hi},v_h X_{hi}^b\right)$ for known $X_{hi}>0$, unknown $\{\alpha_h,v_h\}$, and known $b\ge 0$, and where strata are fixed upfront.
Although the key texts by \citet{Cochran1977} and \citet{Sarndal1992} do not directly provide a solution, \citeauthor{Sarndal1992} address optimal design for a related model under somewhat different problem formulations.
For instance, \citeauthor{Sarndal1992} provide optimal unit-level selection probabilities for the generalized regression estimator (GREG) assuming heteroscedastic errors; separately, they also describe model-based optimal stratification rules.

As implied by Cochran's rules of thumb, populations can differ with respect to the level of heteroscedasticity.
Consider an unstratified superpopulation model of $N$ units, $\{X_i,Y_i:i=1,...,N\}$, where
\begin{align}
	\label{eqn:heteroscedasticity_basic_model_start} &Y_{i} = \beta X_i + \varepsilon_i \\
	&{\mathrm{E}}_\mathrm{M}(\varepsilon_i) = 0 \\
	\label{eqn:heteroscedasticity_basic_model_end} &{\mathrm{Var}}_\mathrm{M}(\varepsilon_i) = \sigma^2 X_i^b 
\end{align}
\noindent for known $X_i>0$, independent $\varepsilon_i$'s, and known $b\ge 0$, and where ${\mathrm{E}}_\mathrm{M}(.)$ and $\mathrm{Var}_\mathrm{M}(.)$ denote the expectation and variance with respect to the model.
If $b=1$, $S_{dh}$ will be roughly proportional to $\sqrt{\bar{X}_h}$, which could justify use of Cochran's allocation of $n_h\propto N_h \sqrt{\bar{X}_h}/\sqrt{c_h}$.
In contrast, $b=2$ could lead to Cochran's allocation of $n_h \propto N_h \bar{X}_h / \sqrt{c_h}$.
As cited by \citet{Henry2009}, $b$ has sometimes been referred to as a `measure of heteroscedasticity' \citep{Foreman1991} or `coefficient of heteroscedasticity' \citep{Brewer2002}.
The above model, as well as a multivariate generalization thereof, is commonly applicable in establishment contexts, with common values of $0\le b \le 2$ \citep[e.g.,][177]{Valliant2000}.

The level of heteroscedasticity has meaningful implications for sample allocation.
Under a probability proportional to size (PPS) design with constant per-unit costs, the optimum unit-level selection probabilities for GREG under this variance structure are $\pi_i \propto X_i^{b/2}$ \citep[e.g.,][sec. 12.2]{Sarndal1992}.
This result, which applies to ratio estimation as a special case, is derived by considering the estimation error $(\hat{Y}-Y)$ for the population total, the latter of which is treated as a random variable (as opposed to the usual design-based treatment of $Y$ as being fixed); \citeauthor{Sarndal1992} define an optimal design as a choice of first-order selection probabilities that minimizes the approximate anticipated variance of $\hat{Y}-Y$ for a given expected sample size, where anticipated variance is defined by \citet{Isaki1982} and is the expected variance with respect to the sampling design and the superpopulation model.
These results can be further generalized by replacing the assumption of Equation (\ref{eqn:heteroscedasticity_basic_model_end}) with the weaker assumption that $\mathrm{Var}_\mathrm{M}(Y_i)=v_i$ for known and positive $v_i$, leading to an optimal allocation of $\pi_i \propto \sigma_i$, a result that \citet[][53]{Valliant2013} attribute to \citet{Godambe1965} and \citet{Isaki1982} for the univariate and multivariate cases, respectively.
These results can also be used to justify equal aggregate stratification rules \citep[e.g.,][189]{Valliant2000}.

Although the model expressed by Equations (\ref{eqn:heteroscedasticity_basic_model_start}--\ref{eqn:heteroscedasticity_basic_model_end}) has many applications, we will subsequently loosen the assumptions to accommodate group differences that do not necessarily depend on size.
Of particular note, the assumption that model variances are of the form $\mathrm{Var}_\mathrm{M}(\varepsilon_i) = \sigma^2 X_i^b$ accommodates heteroscedasticity, but could potentially lead to poor results if the proportionality constant, $\sigma^2$, varies by group.
We also will subsequently allow for slopes to vary by group.
In an establishment survey, the slopes and/or the variance proportionality constant could potentially vary by characteristics other than the size measure, such as industry, firm structure, and/or geography.

\subsection{Sampling with uncertainty}
\label{subsec:background:samp_uncertainty_overview}
There are many works on optimal sample design from a Bayesian perspective, largely published from the mid-1960s through the early 1980s.
This includes the development of optimal STSRS designs for estimating population totals \citep{Draper1968a,Ericson1965,Ericson1967b,Ericson1969b,Ericson1972,Rao1972,Khan1976b} and proportions \citep{Zacks1970,Grosh1972}.
Some of these articles entailed the use of normally distributed data---for instance, \citet{Draper1968a} assumed that the population values within a stratum were normally distributed with separate means and variances, and \citet{Ericson1965} made the somewhat weaker assumption that the strata sample means, conditional on the population means, were normally distributed.
An exception is \citet{Rao1972}, who employed a Dirichlet-multinomial model.
In aiming to minimize the expected posterior loss (e.g., posterior variance), a common approach is to employ a preposterior analysis, which entails computing the posterior distribution (conditional on the data), but then averaging over unknown quantities, since the data have not yet have been collected \citep[e.g.,][]{Draper1968a,Rao1972}.
After this point, optimization methods (e.g., Lagrange multipliers or convex optimization methods) can be used to determine the optimal strata sample sizes.
Although much of the literature focuses on optimization for estimation of a single quantity, there have been some extensions to multivariate problems \citep{Draper1968b,Khan1976,Sekkappan1981b,Sekkappan1981,Soland1967}.
Bayesian optimal allocation methods have also been extended to more complicated sample designs, such as multi-stage sampling \citep{Ericson1973,Ericson1975,Rao1972}, double sampling for nonresponse \citep{Ericson1967,Rao1972,Singh1978,Singh1984}, and multi-phase sampling \citep{Smith1982,Jinn1987}.
Finally, \citet{Ericson1983c,Ericson1983b} provides some results on optimal allocation between two modes wherein one mode is more expensive but provides higher quality measurements.
Summary papers that include some attention to optimal Bayesian sample design are provided by \citet{Solomon1970} and \citet{Ericson1985,Ericson1988}.

Since 1990, there has been little attention to Bayesian optimal sample designs.
Much of the recent work on Bayesian survey statistics is on inferential topics, such as small area estimation (SAE), accounting for complex sample designs, or handling missing data in analyses.
In fact, in examining recent summary articles or texts on Bayesian statistics in the context of surveys \citep[e.g.,][229--230, 259]{Ghosh2009,Rao2011,Gelman2014}, and in conducting literature searches, we did not find mentions of Bayesian methods for optimal sample allocation beyond those cited above, with one exception by \citet{OMalley2016} in the context of SAE.
There has also been recent Bayesian literature on adaptive or responsive survey design, but the focus there has been on developing priors \citep{Coffey2020,West2021} or improving estimates of unknown design parameters \citep{Schouten2018,Wagner2020} rather than directly allocating samples, as we aim to do here.

\subsection{Bayesian decision theory for optimal design}
\label{subsec:background:BayesDecision}
\citet{Wald1950} introduced statistical decision theory, which adapted game theory to the formal analysis of making optimal decisions under uncertainty.
Following \citet{Raiffa1961}, \citet[19--21]{Lindley1972} provided the Bayesian decision theoretic framework on optimal experimental design.
Lindley treats optimal experimental design as a two-part decision: first, the choice of experiment (e.g., sample allocation for an STSRS design), which results in data; second, the choice of how to translate the data into a terminal decision (e.g., compute an estimate), and which, for some given value of the parameter, results in a loss defined by the function $L\left(\hat{\theta},\theta,e,x\right)$.
Under this framework, we are interested in the optimal experiment and estimate that maximize the expected utility, or equivalently, minimize the expected loss.

In a continuous context, the optimal solution that minimizes the expected value of the loss is:
\begin{align}
	\underset{e}{min}\int_X \underset{\hat{\theta}}{min}\int_{\Theta} L\left(\hat{\theta},\theta,e,x\right) p(\theta|x,e) p(x|e) \, d\theta \, dx
\label{eqn:Lindley1}
\end{align}
for loss function $L\left(\hat{\theta},\theta,e,x\right)$, experiment $e$, and the sample space $X$ modeled by $\theta\in \Theta$ and estimated by $\hat{\theta}(x) \equiv \hat{\theta}$ for sample $x \in X$. 
The distribution $p(x|e)$ is a conditional pdf for obtaining a particular sample given the experiment, while $p(\theta|x,e)$ is a posterior pdf for the population parameter given the sample and experiment.
The inner minimization step, termed the \emph{terminal analysis}, averages the loss function over $\Theta$ (treating the sample as fixed) then optimizes $\hat{\theta}$.
The outer minimization, termed the \emph{preposterior analysis}, averages the inner minimization over the predictive distribution for the data then optimizes $e$.

In this manuscript we assume squared error loss (SEL): $L\left(\hat{\theta},\theta,e,x\right)=\left(\hat{\theta}-\theta\right)^2$. Then (\ref{eqn:Lindley1}) becomes
\begin{align}
	\underset{e}{min}\int_X \underset{\hat{\theta}}{min}\int_{\Theta} L\left(\hat{\theta},\theta,e,x\right) p(\theta|x,e) p(x|e) \, d\theta \, dx
 & = 
	\underset{e}{min}\int_X \underset{\hat{\theta}}{min} \,  \underset{\theta|x,e}{\mathrm{E}}(\hat{\theta}-\theta)^2 p(x|e) \, dx \\ \nonumber
 & = 
	\underset{e}{min}\int_X  \mathrm{Var}(\theta|x,e) p(x|e) \, dx \\ \nonumber
 & = 
\underset{e}{min}\left\{\underset{x|e}{\mathrm{E}}\mathrm{Var}(\theta|x,e)\right\}
\end{align}
Thus, under SEL, the preposterior analysis entails considering $\underset{x|e}{\mathrm{E}}\mathrm{Var}(\theta|x,e)$, which is the expected posterior variance, and then selecting $e$ in order to minimize this quantity.

Hence, we can implement the following algorithm to identify the optimal design:
\begin{enumerate}
\item Specify $L\left(\hat{\theta},\theta,e,x\right)$ (we assume SEL), $\theta$ (finite population mean), $e$ (sample allocation to strata).
  \item Find the estimate that minimizes the expected loss (posterior of mean under SEL).
	\item Find the posterior loss ($V(\theta|x,e)$ in our setting).
	\item Conduct preposterior analysis: average over the predictive distribution for the main study data given the pilot data to obtain the expected posterior loss for a given sample size. This requires deriving an approximate closed form for the expected posterior loss for a given allocation via a first-order Taylor series approximation.
	\item Find the optimal allocation for minimizing the expected posterior loss by onsidering the space of possible sample allocations using standard numerical methods.
\end{enumerate}
In the next section, we provide the analytical results required to implement this algorithm for stratified random sampling for our specific heteroscedastic model of interest.

\section{Analytical results}
\label{sec:ch2:BayesSTSRS_analytic_results}

\subsection{Notation}
Strata are given by $h=1,...,H$. 
The finite population total and mean within stratum $h$ are given by $Y_h = \sum_{i=1}^{N_h}Y_{hi}$, and $\bar{Y}_h=\frac{Y_h}{N_h}$, respectively; similarly the finite population total and mean for the auxiliary variable are given by 
$X_h=\sum_{i=1}^{N_h}X_{hi}$ and $\bar{X}_h=\frac{X_h}{N_h}$. The target parameter of interest is the population total
$\theta=\bar{Y} = \frac{1}{N} \sum_{h=1}^H  Y_h $, where $N = \sum_{h=1}^H N_h$.

$D1$ and $D2$ refer to the pilot and main study data respectively and $m_h$ and $n_h$ refer to the sample sizes in stratum $h$ for $D1$ and $D2$.
Define the sample allocation to the strata as
$e_1=\left\{n_h:h=1,...,H\right\}$ under the constraint that, e.g., $\sum_{h=1}^H n_h =n$ for prespecified $n$ or $\sum_{h=1}^H n_h c_h \le C$ for a constraint on total cost $C$, and sample size constraints of the form $4\le n_h \le N_h$.
The sets of units sampled in stratum $h$ in the pilot and main study are denoted by $s_{1h}$ and $s_{2h}$, respectively, and $s_{1h}^C$ and $s_{2h}^C$ are the complements of these sets over the set of stratum population units $U_h$.
The sample means of the outcome and auxiliary variables for the main study are given by $\bar{y}_h=\frac{1}{n_h}\sum_{i\in s_{2h}}y_{hi}$ and $\bar{x}_h=\frac{1}{n_h}\sum_{i\in s_{2h}}x_{hi}$. 
Finally we denote the transformed sampled auxiliary variable by $z_{hi}=x^b_{hi}$.

\subsection{Heteroscedastic error model}
We consider a population model wherein each stratum has its own slope, and the variance of
a stratum’s error term is proportional to a fixed power of an auxiliary variable. Thus our model of interest is
\begin{align}
Y_{hi}=\alpha_h X_{hi}+\varepsilon_{hi}
\end{align}
where $\varepsilon_{hi}\overset{ind}{\sim} N(0,v_h X_{hi}^b)$ and $X_{hi}>0$ for stratum $h=1,...,H$ with population units $i=1,...,N_h$. Assume $H$ strata are predefined, and that each phase uses stratified random sampling (STSRS).
We treat $\{\alpha_h,v_h\}$ are unknown and assume that $b\ge 0$ and auxiliary variable $\{X_{hi}\}$ are known
for the population. 
Transforming our scale parameter $g_h = \frac{1}{v_h}$, we further assume a non-informative prior on $(\alpha_h, g_h)$ given by $\pi(\alpha_h,\frac{1}{v_h}) \propto v_h$. This prior leads to a normal-gamma posterior for $(\alpha_h,\frac{1}{v_h})$.

We assume the pilot data are used only for planning the main study, and are not to be used in our ultimate inferences about the population mean. However, our results could be easily extended to allow for a different decision on this matter.

\subsection{Find the optimal estimator}
As per \citet{Lindley1972}, the first step to finding the Bayesian optimal design is to specify a loss function as a function of the parameter (i.e., finite population mean), estimate, experiment (i.e., sample allocation), and sample, after which we find the estimator that minimizes the expected loss conditional on the data.
We specify a loss function of $L(\bar{Y},\hat{\bar{Y}},e_1,D2)=\left(\hat{\bar{Y}}-\bar{Y}\right)^2$.
As per our Bayesian formulation, $\bar{Y}$ is a parameter, and is random with respect to the parameter space.
Let $\hat{\bar{Y}}\left(D2\right)$ denote some estimator for the finite population mean as a function of the main study data (D2).

Holding the main sample fixed leads to an expected loss of $\underset{\bar{Y}|D2}{\mathrm{E}}\left\{\left(\bar{Y}-\hat{\bar{Y}}(D2)\right)^2 | D2 \right\} =  \mathrm{Var}(\bar{Y}|D2) + \left(\mathrm{E}(\bar{Y}|D2)-\hat{\bar{Y}}(D2)\right)^2$.
This expression is minimized when estimation is via the posterior mean, leading to an expected loss of $\underset{\bar{Y}|D2}{\mathrm{E}}\left(\bar{Y}-\hat{\bar{Y}}\right)^2=\mathrm{Var}\left(\bar{Y}|D2\right)$, which is the posterior variance for the finite population mean.
Thus, we will ultimately specify our optimization problem as selecting the sample size to minimize the expected posterior variance, from within the space of feasible sample allocations.

\subsection{Find the posterior loss}
\label{subsec:fixed_b_posterior_loss}
Next, we find the posterior loss, which is the posterior variance for the finite population mean, and which we write as $\mathrm{Var}\left(\bar{Y}|D2,e_1,b\right)$ as to emphasize that we are fixing not only the data but also the sample allocation and coefficient of heteroscedasticity.
We can find the posterior loss through an application of results from \citet[][sec. 5.1]{Ericson1969} for each stratum.
For stratum $h$, for $h=1,...,H$, we assume a diffuse prior on variables $\left(\alpha_h,\frac{1}{v_h}\right)$ with density $\pi\left(\alpha_h,\frac{1}{v_h}\right)\propto v_h$, as per Appendix A.2, %\ref{app:EricsonPosteriorSection},
and we also assume that the $H$ pairs of variables are independent across strata.
Then, we can compute $\mathrm{E}\left(\bar{Y}_h | D2,e_1,b\right)$ and $\mathrm{Var}\left(\bar{Y}_h | D2,e_1,b\right)$ through a direct application of Ericson's Equations (73) and (74), after which we combine results across strata.
Assuming that $n_h > 3$ for all $h$, this results in:

\begin{footnotesize}
\begin{align}
	\mathrm{E}\left(\bar{Y} | D2,e_1,b\right)  &= \sum_{h=1}^H \frac{1}{N} \left\{n_h \bar{y}_h + \left(N_h\bar{X}_h-n_h \bar{x}_h\right)\left(\frac{\sum_{i\in s_{2h}} \frac{x_{hi}y_{hi}}{z_{hi}}}{\sum_{i\in s_{2h}}\frac{x_{hi}^{2}}{z_{hi}}}\right)\right\} \label{eqn:eqn3}
	\\ \mathrm{Var}\left(\bar{Y} | D2,e_1,b\right) &= \sum_{h=1}^H \frac{1}{N^2}\left\{\frac{1}{n_h-3} \left[\sum_{i\in s_{2h}}\frac{y_{hi}^2}{z_{hi}} - \frac{\left(\sum_{i\in s_{2h}}\frac{x_{hi}y_{hi}}{z_{hi}}\right)^2}{\sum_{i\in s_{2h}}\frac{x_{hi}^2}{z_{hi}}} \right] \left[\sum_{i \in (U_h \cap s_{2h}^C)} z_{hi} + \frac{\left(N_h\bar{X}_h-n_h\bar{x}_h\right)^2}{\sum_{i\in s_{2h}} \frac{x_{hi}^2}{z_{hi}}}\right]\right\} \label{eqn:Var_mainstudy}
\end{align}
\end{footnotesize}

\noindent 
(See Appendix A.3 %\S\ref{app:posterior_finitepopmean} 
for detail.)

\subsection{Conduct the preposterior analysis}
\label{subsec:Known_b_preposterior}
The posterior variance from Equation (\ref{eqn:Var_mainstudy}) above is a function of known population quantities, the main study sample sizes, and the main study sample data.
However, the main study sample data have not yet been collected.
Therefore, as per \citeauthor{Lindley1972}, the next step is a preposterior analysis, which entails averaging over the predictive distribution $D2|D1$.
Specifically, we need to compute $\underset{D2|D1}{\mathrm{E}}\left\{\mathrm{Var}\left(\bar{Y}|D2,e_1,b\right)\right\}$, which is the expected posterior loss for a given sample allocation.

%app:fixed_known_b_analysis
%Analysis for fixed and known b
Assuming large $\{n_h\}$, we show in Appendix B %\ref{app:fixed_known_b_analysis} that 
that
\begin{align}
\underset{D2|D1}{\mathrm{E}} \mathrm{Var}\left(\bar{Y} | D2,e_1,b\right) \approx \sum_{h=1}^H \frac{\mathrm{E}_P (v_h)}{N^2} \left(\frac{n_h-1}{n_h-3}\right) \left(N_h-n_h\right) \left[\bar{Z}_h+ \frac{\frac{n_h}{N_h} S^2_{xh} + (N_h-n_h) \bar{X}_h^2}{n_h \bar{Q}_h^2 (1 + CV_{qh}^2)}\right] \label{eqn:EV_fixed_b_final}
\end{align}
\noindent where selected terms above are defined as $\mathrm{E}_P(v_h)=\underset{\alpha_h,v_h|D1}{\mathrm{E}}(v_h)$, $S^2_{xh} = \frac{1}{N_h-1} \sum_{i=1}^{N_h} (X_{hi}-\bar{X}_h)^2$, $Q_{hi}=\frac{X_{hi}}{\sqrt{Z_{hi}}}$, $\bar{Q}_h=\frac{1}{N_h}\sum_{i=1}^{N_h} Q_{hi}$, $CV_{qh} = \frac{S_{qh}}{\bar{Q}_h}$, and $S^2_{qh}=\frac{1}{N_h-1}\sum_{i=1}^{N_h}\left(Q_{hi}-\bar{Q}_h\right)^2$, and where it turns out that $\mathrm{E}_P(v_h) = \frac{1}{m_h-3} \left(\sum_{i\in s_{1h}}\frac{y_{hi}^2}{z_{hi}}-\frac{\left(\sum_{i\in s_{1h} }\frac{y_{hi} x_{hi}}{z_{hi}}\right)^2}{\sum_{i\in s_{1h}}\frac{x_{hi}^2}{z_{hi}}}\right)$.
The exact result, which does not require large $\{n_h\}$, is $\underset{D2|D1}{\mathrm{E}} \mathrm{Var}\left(\bar{Y} | D2,e_1,b\right) = \sum_{h=1}^H \frac{1}{N^2} \left(\frac{n_h-1}{n_h-3}\right) \left[\mathrm{E}_P (v_h)\right] \mathrm{E}\left(k(s_{2h})\right)$, where $k(s_{2h})=\left[\sum_{i \in (U_h \cap s_{2h}^C)} z_{hi} + \frac{\left(N_h\bar{X}_h-n_h\bar{x}_h\right)^2}{\sum_{i\in s_{2h}} \frac{x_{hi}^{2}}{z_{hi}}}\right]$.
The approximate result shown in Equation (\ref{eqn:EV_fixed_b_final}) assumes large $\{n_h\}$ in applying
a first-order Taylor series (TS) approximation for $\mathrm{E}\left(k(s_{2h})\right)$.

The TS approximation above was used on account of the rightmost term in $k(s_{2h})$, which is nonlinear.
Alternatively, since $k(s_{2h})$ depends solely on the sample indicators and auxiliary variable, one could compute a Monte Carlo (MC) estimate of $\mathrm{E}\left(k(s_{2h})\right)$, as a function of $\{X_{hi}\}$ and $b$, by averaging over $R$ design-based samples, for large $R$.
If the MC estimator is directly incorporated into the objective function, it may be helpful to construct a high-quality and quickly-evaluated MC emulation function for $\hat{\mathrm{E}}\left(k(s_{2h})|\left\{n_h\right\}\right)$ for given $\{n_h\}$, as to avoid computational bottlenecks during the optimization process; we describe how to do so in Appendix B.4.%\ref{app:MC_approx_for_ks2h}.

\subsection{Obtain optimal allocation}
\label{subsec:Mendelson_optimal_alloc_fixed_b}
Equation (\ref{eqn:EV_fixed_b_final}) above provides a closed form expression for the approximate expected posterior loss as a function of known population quantities, the pilot sample, and the main study sample allocation.
Hence, the optimal allocation, $\underset{e_1}{argmin} \underset{D2|D1}{\mathrm{E}} \mathrm{Var}\left(\bar{Y} | D2,e_1,b\right)$, can be obtained via standard mathematical programming methods \citep[see, e.g.,][chap. 5]{Valliant2013}.
A common formulation would entail aiming to minimize Equation (\ref{eqn:EV_fixed_b_final}) subject to a constraint on total costs, $\sum_{h=1}^H n_h c_h \le C$, and sample size constraints of the form $4\le n_h \le N_h$.
Here, we have assumed that $n_h\ge 4$ on account of the $(n_h-3)^{-1}$ term in Equation (\ref{eqn:Var_mainstudy}).
Survey designers could also add any other relevant constraints (e.g., precision constraints for domain estimation).

After computing an allocation, if Equation (\ref{eqn:EV_fixed_b_final}) was used, it may be helpful to use MC sampling to assess the appropriateness of the Taylor series approximation for $\mathrm{E}\left(k(s_{2h})\right)$, as suggested in the last paragraph of \S\ref{subsec:Known_b_preposterior}.

\section{Simulation study}

\label{sec:ch2:Simulation_design}

\subsection{Overview}

Simulation methods were used to compare performance of the proposed sampling strategy with alternative stratified random sampling (STSRS) designs.
For a given population, each simulation entailed drawing a pilot sample and then, for each sampling strategy, computing the sample allocation, drawing a main study sample, and making inferences according to that strategy's estimation methods. Although not limited to establishment surveys, such surveys are a likely setting for implementing these methods.  Unfortunately, due to confidentiality-related issues, there is a lack of publicly available establishment survey microdata.  Therefore, we compared our proposed sample allocation method with alternatives across a series of artificially generated populations that aimed to reflect selected characteristics of establishment surveys. Specifically, we created $P=90$ bivariate populations, 
consisting of
three shapes for the auxiliary variable crossed with five coefficients of heteroscedasticity, further crossed with six structures for the slope and correlation for the two variables.
We considered a single mechanism for stratification.
Then, we compared results primarily in terms of RMSE, relative bias, and coverage and relative width of 95\% CIs.

\subsection{Data generating mechanism}
\label{subsec:Simulation:popdata}

We generated the covariate $X$ for our population by first generating 12,500 observation from a $GAMMA(k,\theta)$ distribution
and then dropping the first and last deciles, resulting in truncated gamma distributions of $N=10,000$ units. The gamma distribution is relevant to establishment data, which are continuous and heavily skewed; thus we can consider $X$ as a measure of size (MOS).
Truncation was done to reduce the numbers of near-zero and extremely large units, the latter of which would presumably be in a take-all stratum under any design and wherein practical considerations would commonly hinge upon reducing non-sampling errors. We considered three different gamma distributions:
\begin{itemize}
\item $(k=0.2,\theta=5)$: MOS 1, highly skewed
\item $(k=0.6, \theta=5/3)$: MOS 2, moderately skewed
\item $(k=1, \theta=1)$: MOS 3, slightly skewed
\end{itemize}

\begin{figure}[h]
\centering
\caption{Distributions of simulated stratified size measures, by MOS}
\label{fig:sim_stratified_size_measures}
\includegraphics{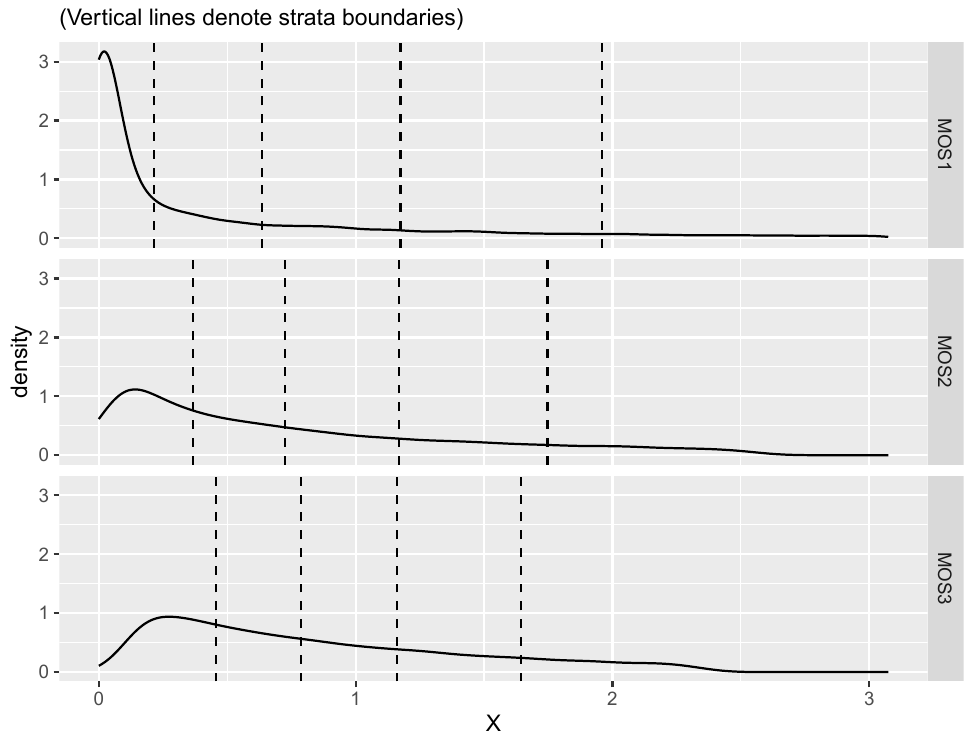}
\end{figure}

Next, five strata were then formed based on equalizing the aggregated square roots of $X_i$. This was operationalized by sorting the $\{X_i:i=1,...N\}$ in ascending order, computing the cumulative sum of the first $k$ elements as $c(k) = \sum_{i=1}^{k}\sqrt{X_i}$, for $k=1,...,N$, and then using cut points of $\frac{1}{5}\sum_{i=1}^{N}\sqrt{X_i}$, $\frac{2}{5}\sum_{i=1}^{N}\sqrt{X_i}$, $\frac{3}{5}\sum_{i=1}^{N}\sqrt{X_i}$, and $\frac{4}{5}\sum_{i=1}^{N}\sqrt{X_i}$ to group units based on the cumulative sums. Relating our size measures and stratification to what is done in practice, establishment surveys commonly stratify by industry and/or unit size; for instance, several of the Census Bureau's economic surveys have stratified by size within industry \citep{NAP2018}.
The resulting univariate distributions we generated are displayed in Figure \ref{fig:sim_stratified_size_measures}.

%C:/Users/jmend/Documents/JPSM/Dissertation/Data analysis/Paper 1 replication attempt/analysis/Simulation_data_viz_v6f_supplement.rmd
% NOTE: graphic generated via replication of paper 1 original plots

%\subsubsection{Model parameters}
%\label{ch2_subsubsec_modelparams}

\begin{figure}[p]
	\centering
	\caption{Distributions of simulated stratified size measures: MOS1, b=1 populations}
	\label{fig:sim_MOS1_b1_bivariate_dists} 
	\includegraphics[scale=1]{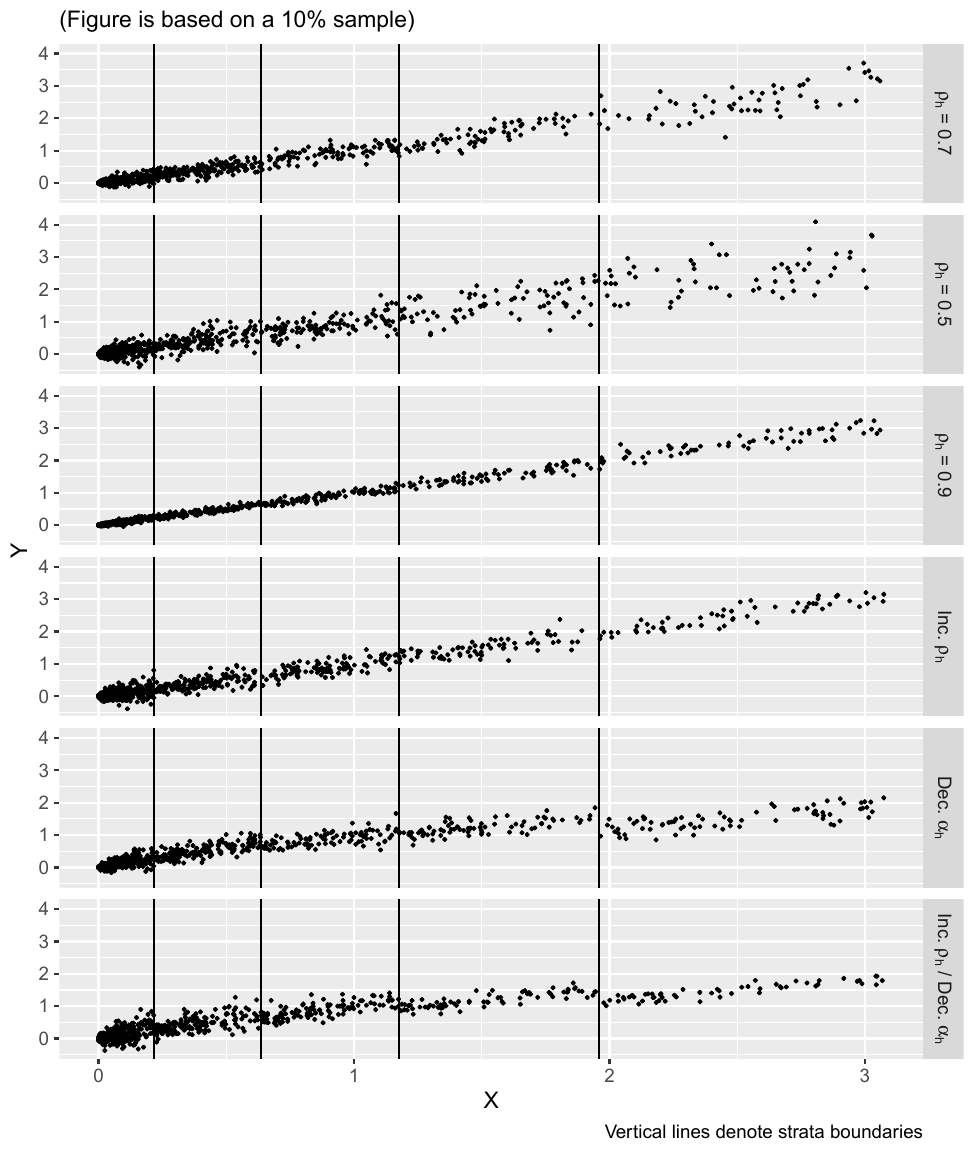}
\end{figure}

Finally, we generated the outcome variable as $Y_{hi}=\alpha_h X_{hi}+\varepsilon_{hi}$
where $\varepsilon_{hi}\overset{ind}{\sim} N(0,v_h X_{hi}^b)$ for stratum $h=1,...,5$ with population units $i=1,...,N_h$.  We considered $b= 0$, $b=0.5$, $b=1$, $b=1.5$, and $b=2$, consistent with the coefficient of heteroscedasticity typically lying within the interval $[0,2]$ in real applications.  For each value of $b$, we then developed six structures for the remaining parameters, $\{\alpha_h,v_h\}$. We specified a set of slopes, $\{\alpha_h\}$, and a set of target correlations, $\{\rho_h\}$, and solved for $v_h|\alpha_h,\rho_h,b$, such that the correlation between the $\{X_{hi},Y_{hi}\}$ for a given stratum was approximately equal to $\rho_h$:
\begin{align}
v_h = \frac{\alpha_h^2 S^2_{xh}\left(\frac{1}{\rho^2_h}-1\right)}{\bar{Z}_h}, \label{eqn:vh_given_rhoh}
\end{align}
where $\bar{Z}_h=\frac{1}{N_h}\sum_{i=1}^{N_h}Z_{hi}$, $Z_{hi}=X_{hi}^b$, $S^2_{xh}=\frac{1}{N_h-1}\sum_{i=1}^{N_h}\left(X_{hi}-\bar{X}_h\right)^2$, and  $\bar{X}_h=\frac{1}{N_h}\sum_{i=1}^{N_h}X_{hi}$ (see Appendix C.1 %\ref{app:paper1_target_correlation_explanation} 
for detail).
%app:paper1_target_correlation_explanation
%Selection of \texorpdfstring{$v_h$}{v\_h} to achieve target correlation
The six scenarios specified were as follows:
\begin{enumerate}
\item \textbf{Baseline}: assume correlations of $\rho_1=\rho_2=\rho_3=\rho_4=\rho_5=0.7$ and slopes of $\alpha_1=\alpha_2=\alpha_3=\alpha_4=\alpha_5=1$, where strata 1 through 5 are ordered based on ascending size measures (e.g., stratum 1 refers to the units with the smallest X's).
\item \textbf{Lower correlations}: assume correlations of $\rho_1=\rho_2=\rho_3=\rho_4=\rho_5=0.5$ and slopes of $\alpha_1=\alpha_2=\alpha_3=\alpha_4=\alpha_5=1$.
\item \textbf{Higher correlations}: assume correlations of $\rho_1=\rho_2=\rho_3=\rho_4=\rho_5=0.9$ and slopes of $\alpha_1=\alpha_2=\alpha_3=\alpha_4=\alpha_5=1$.
\item \textbf{Increasing correlations, fixed slopes}: assume monotonically increasing correlations of $\rho_1=0.5$, $\rho_2=0.6$, $\rho_3=0.7$, $\rho_4=0.8$, and $\rho_5=0.9$ and constant slopes of $\alpha_1=\alpha_2=\alpha_3=\alpha_4=\alpha_5=1$.
\item \textbf{Fixed correlations, decreasing slopes}: assume correlations of $\rho_1=\rho_2=\rho_3=\rho_4=\rho_5=0.7$ and monotonically decreasing slopes of $\alpha_1 = 1.4$, $\alpha_2=1.2$, $\alpha_3=1.0$, $\alpha_4=0.8$, and $\alpha_5=0.6$.
\item \textbf{Increasing correlations, decreasing slopes}: assume monotonically increasing correlations of $\rho_1=0.5$, $\rho_2=0.6$, $\rho_3=0.7$, $\rho_4=0.8$, and $\rho_5=0.9$, and monotonically decreasing slopes of $\alpha_1 = 1.4$, $\alpha_2=1.2$, $\alpha_3=1.0$, $\alpha_4=0.8$, and $\alpha_5=0.6$.
\end{enumerate}
The first three scenarios assumed equal slopes and correlations across strata, with slopes fixed at 1 and correlations fixed at 0.5, 0.7, or 0.9.
The last three scenarios allowed the correlations to increase and/or slopes to decrease with respect to increasing size measures, which could possibly be motivated by the up-or-out dynamic of young firms.
On the latter point, \citet{Haltiwanger2013} observed a strong inverse relationship between firm size and net employment growth using microdata from the Census Bureau's Longitudinal Business Database; the authors attributed this relationship to the role of young firms, which tended to be smaller and more volatile than older firms.
Thus, our latter scenarios allowed for strata with larger units to have larger correlations between the X's and Y's (e.g., larger correlation between previous and current year's revenue) and/or smaller slopes (e.g., lower revenue growth). Figure \ref{fig:sim_MOS1_b1_bivariate_dists}
displays six bivariate populations that used MOS1 and where $b=1$, for a randomly selected ten-percent sample of each population; each panel reflects a different slope/correlation scenario.
%fig:sim_MOS1_b1_bivariate_dists
%Fig name. Distributions of simulated stratified size measures: MOS1, b=1 populations
%Appendix name. Supplementary figures and tables

%NOTE: changed from "H" to "p" to get a reasonable placement ("h" was too far ahead).
%Could change positioning again if I reduce the figure to fit as portrait

\subsection{Pilot design}
\label{subsec:Simulation:pilotdata}

An equally allocated pilot sample $D1$ of size 75 was drawn for each simulation, so that $m_1=\cdots=m_5=15$; simple random sampling was conducted within each stratum. Although, in practice, the allocation for a pilot sample will depend on the specific study, equal allocation can be justified in some situations, for instance, in scenarios where a sample designer has no reason, a priori, to prioritize certain strata.

\subsection{Main study allocations and estimators}
\label{subsec:Simulation:mainstudy}
Next, we applied competing strategies for each given pilot data set, wherein a strategy comprises an allocation method and an estimator.
For the $p$th population and $r$th simulation, we used the pilot data set $d_1^{(p,r)}$ to compute $A$ different allocations (i.e., one per strategy), then drew a single design-based sample for each allocation, and computed the estimated population total along with variance estimates and associated 95\% equal-tail CI (confidence interval or credible interval) for each sample.
We fixed the total sample size at $n=500$ units.
A total of 1000 simulations were generated to evaluate each data generating mechanism.

For our B-P (Bayesian allocation/prediction estimation) strategy, the sample allocation was computed to minimize the expected posterior loss subject to constraints of $4 \le n_h \le N_h$.
The expected posterior loss, expressed in Equation ($\ref{eqn:EV_fixed_b_final}$), incorporates a TS approximation for $\mathrm{E}(k_{s_{2h}})$; as an alternative, we computed Monte Carlo (MC) estimates of $\mathrm{E}(k_{s_{2h}})$, using the methods described in Appendix B.4, and based on 10,000 random permutations of the population.
%app:MC_approx_for_ks2h. {Sampling-based approximation for \texorpdfstring{$k(s_{2h})$}{k(s2h)}}
Sample optimization used a COBYLA-based algorithm \citep{Powell1994}, as implemented in \textsf{NLopt} \citep{Johnson2014}, and inferences were made using the prediction estimator from Equation ($\ref{eqn:eqn3}$).
Given that the MC- and TS-based implementations of B-P produced nearly identical results, we report on the TS-based methods, which are simpler to implement and are reasonable to use for large $\{n_h\}$.

In addition to B-P, we consider five alternative stratified random sampling (STSRS) designs, assuming constant per-unit costs.
\begin{enumerate}
\item The fully design-based Neyman allocation, in conjunction with the HT-estimator (i.e., N-HT strategy).

\item \citeauthor{Cochran1977}'s model-assisted strategy for separate ratio (SR) estimation (C-SR strategy; see \S\ref{subsec:Cochran_allocation_separate_ratio_estimator}).
The allocation is $n_h \propto N_h \hat{S}_{dh}$, where $\hat{S}_{dh}$ given by Equation (\ref{eqn:Cochran_plugin_SS_estimator}) is a plug-in estimator for $S_{dh}$, the standard deviation of a residual term from a SR model; inferences are by the SR estimator.% (Equation [\ref{eqn_SR_estimator}]).
\item A Cochran ``rule of thumb'' variation of 2 (RT0-SR)
given by $n_h \propto N_h$.
\item A Cochran ``rule of thumb'' variation of 2 (RT1-SR) given by $n_h\propto N_h \sqrt{\bar{X}_h}$.
\item A Cochran ``rule of thumb'' variation of 2 (RT2-SR) given by $n_h\propto N_h\bar{X}_h$.
\end{enumerate}
Strategies 3--5 may sometimes be appropriate when $b=0$, $b=1$, or $b=2$, respectively, and when the $v_h$ terms are constant across strata.	

For all strategies, sample allocations were rounded, incorporating adjustments when necessary to ensure total sample sizes of 500.
The latter involved multiplying a given allocation by $\left(1+\epsilon^{(r)}_{(p,a)}\right)$, before rounding, for some $\epsilon^{(r)}_{(p,a)}$ close to zero that was chosen to ensure the intended total sample size.

The above strategies are summarized in Table \ref{table:overview_of_sim_strategies}. 

\begin{singlespace}
		\begin{table}[!h]
\centering\centering
\caption{\label{tab:unnamed-chunk-2}Overview of main sampling strategies for simulation}
\centering
\fontsize{9}{11}\selectfont
\begin{tabular}[t]{>{\raggedright\arraybackslash}p{1.15cm}>{\raggedright\arraybackslash}p{2.91cm}>{\raggedright\arraybackslash}p{2.19cm}>{\raggedright\arraybackslash}p{2.4cm}>{\raggedright\arraybackslash}p{2.02cm}>{\raggedright\arraybackslash}p{3.17cm}}
\toprule
 & Strategy & Inferential approach & Allocation & Estimator & Comments\\
\midrule
\cellcolor{gray!10}{N-HT} & \cellcolor{gray!10}{Neyman/HT} & \cellcolor{gray!10}{Design-based} & \cellcolor{gray!10}{Neyman with assumption of $S_h = \hat{S}_h$} & \cellcolor{gray!10}{HT estimator} & \cellcolor{gray!10}{}\\
\addlinespace
C-SR & Cochran plug-in/ SR estimator & Design-based/ model-assisted & Cochran (1977), $\S$6.14, with assumption of $S_{dh} = \hat{S}_{dh}$ & SR estimator & $\hat{S}_{dh}$ not directly provided by Cochran\\
\addlinespace
\cellcolor{gray!10}{RT0-SR} & \cellcolor{gray!10}{Proportional allocation/ SR estimator} & \cellcolor{gray!10}{Design-based/ model-assisted} & \cellcolor{gray!10}{$n_h \propto N_h$} & \cellcolor{gray!10}{SR estimator} & \cellcolor{gray!10}{Implied by Cochran if $S_{dh}$ is roughly constant}\\
\addlinespace
RT1-SR & $b=1$ rule of thumb/ SR estimator & Design-based/ model-assisted & $n_h \propto N_h\sqrt{\bar{X}_h}$ & SR estimator & Suggested by Cochran if $S_{dh}$ is roughly proportional to $\sqrt{\bar{X}_h}$\\
\addlinespace
\cellcolor{gray!10}{RT2-SR} & \cellcolor{gray!10}{$b=2$ rule of thumb/ SR estimator} & \cellcolor{gray!10}{Design-based/ model-assisted} & \cellcolor{gray!10}{$n_h \propto N_h \bar{X}_h$} & \cellcolor{gray!10}{SR estimator} & \cellcolor{gray!10}{Suggested by Cochran if $S_{dh}$ is roughly proportional to $\bar{X}_h^2$}\\
\addlinespace
B-P & Bayesian allocation/ prediction estimator & Bayesian & Minimize Eqn. ($\ref{eqn:EV_fixed_b_final})$ assuming that $b = b^{(p)}$ & $\mathrm{E}(Y|D2,e_1,b)$ via Eqn. ($\ref{eqn:eqn3}$) and $b = b^{(p)}$ & \\
\bottomrule
\end{tabular}
\end{table}

		{\makeatletter\def\@currentlabel{\thetable}\label{table:overview_of_sim_strategies}}
	\end{singlespace}

\subsubsection{Variance estimation}
We used classical variance estimators for the HT and SR estimators.
HT variances were estimated via $\mathrm{var}(\hat{Y})=\sum_{h=1}^H\frac{N_h^2}{n_h}\left(1-\frac{n_h}{N_h}\right)\frac{1}{n_h-1}\sum_{i=1}^{n_h}{(y_{hi}-\bar{y}_h)^2}$.
SR variances were estimated via $v_2(\hat{Y}_{Rs}) = \sum_{h=1}^{H} \left(\frac{\bar{X}_h}{\bar{x}_h}\right)^2 \frac{N_h^2(1-f_h)}{n_h}\frac{1}{n_h-1}\sum_{i=1}^{n_h} (y_{hi} - \hat{R}_h x_{hi})^2$, where $\hat{R}_h = \frac{\bar{y}_h}{\bar{x}_h}$ is the ratio of sample means in stratum $h$.
For B-P, the posterior variance was computed via Equation (\ref{eqn:Var_mainstudy}).

\subsection{Performance comparison}
\label{subsec:Simulation:performance_comparison}

Using the above procedures, let $\hat{Y}_{(p,a)}$ denote the estimator for population $p$ and estimation strategy $a$, and let $\hat{Y}_{(p,a)}^{(r)}$ denote the corresponding estimate for the $r$th simulation, for $p=1,...,P$, $a=1,...,A$, and $r=1,...,R$.
Let $Y_{(p)}$ denote the true population total for population $p$.
We computed four key metrics for a given population $p$ and strategy $a$:
\begin{enumerate}
\item \textbf{RMSE}.
Root mean square error was estimated as \\$rmse\left(\hat{Y}_{(p,a)}\right) = \sqrt{\frac{1}{R} \sum_{r=1}^{R} \left(\hat{Y}^{(r)}_{(p,a)} - Y_{(p)}\right)^2}$.
\item \textbf{Relative bias.}
Bias was estimated as $bias\left(\hat{Y}_{(p,a)}\right) = \frac{1}{R} \left(\sum_{r=1}^{R} \hat{Y}^{(r)}_{(p,a)}\right) - Y_{(p)}$ and relative bias was estimated as $relbias(\hat{Y}_{(p,a)}) = \frac{bias(\hat{Y}_{(p,a)})}{Y_{(p)}}$.
\item \textbf{Coverage of 95\% CIs}.
Compute a 95\% CI for each $\hat{Y}^{(r)}_{(p,a)}$. 
Let $I^{(r)_{(p,a)}}$ be an indicator that is 1 if the CI contains $Y_{(p)}$ and 0 otherwise.
Coverage was estimated as $coverage(p,a)=\frac{1}{R}\sum_{r=1}^{R}I^{(r)}_{(p,a)}$.
\item \textbf{Relative width of 95\% CIs}.
Estimate the average width of a 95\% CI via $width(p,a)=\frac{1}{R}\sum_{r=1}^{R}W^{(r)}_{(p,a)}$, where $W^{(r)}_{(p,a)}$ is the width of the 95\% CI for population $p$, simulation $r$, strategy $a$.
The estimated average relative width is $relwidth(p,a) = \frac{width(p,a)}{Y_{(p)}}$.
\end{enumerate}

For each metric, we also computed Monte Carlo error, defined as the standard deviation of a simulation estimator with respect to repeated implementations of the simulation \citep[e.g.,][]{Koehler2009}.
This error was estimated via $sd(\hat{\theta})=\sqrt{\frac{1}{R(R-1)}
\sum_{r=1}^{R}\left(\hat{\theta}_r-\hat{\theta}\right)^2}$, 
where  $\hat{\theta}_r$ is an estimate obtained from simulation $r$, and $\hat{\theta}=\frac{1}{R}\sum_{r=1}^{R}\hat{\theta}_r$.

For allocation methods that incorporated pilot sample information, we also considered the distribution of sample allocations with respect to repeated pilot sampling.
For population $p$ and strategy $a$, we computed the average allocation for stratum $h$ as
\begin{align}
\bar{n}_{h(p,a)} = \frac{1}{R}\sum_{r=1}^{R}n_{h(p,a)}^{(r)}, \label{eqn:sim_design_avg_alloc}
\end{align}
where $n_{h(p,a)}^{(r)}$ is the stratum $h$ allocation for the $r$th simulation.
Likewise, we computed the standard deviation via
\begin{align}
{sd}_{h(p,a)}  = \sqrt{\frac{1}{R}\sum_{r=1}^{R}\left(n_{h(p,a)}^{(r)}-\bar{n}_{h(p,a)}\right)^2}. \label{eqn:sim_design_sd_alloc}
\end{align}

\subsection{Sensitivity analysis for misspecification of heteroscedasticity}
Although the B-P strategy assumes that the sample designer knows the heteroscedasticity measure $b$ upfront and uses it to design an efficient sample, this may not always be realistic.
Therefore, we conducted a sensitivity analysis to examine the implications of misspecification of $b$ at the sample allocation stage.
For each population, we evaluated four variations of B-P, wherein $b$ was misspecified as 0, 0.5, 1, 1.5, or 2 at the sample allocation stage, but where the true value was used for data generation.

\section{Simulation results}
\label{sec:ch2:sim_results}

\subsection{Main strategies}

First, we consider the N-HT, C-SR, and B-P strategies, which are potentially applicable for any level of $b$.
All three methods were approximately unbiased across populations (Appendix C.2, Table 2), %\ref{tab:RelBias_by_metric}), 
with empirical relative bias that was either indistinguishable from zero or not of practical significance.
%tab:RelBias_by_metric
%Appendix table: %1,000*RelBias by main strategy and population-generating characteristics
Further, all three achieved near-nominal coverage for 95\% CIs (Appendix C.2, Table 3), %\ref{tab:Coverage_by_metric}), 
with any deviations appearing to be attributable to Monte Carlo error.
%tab:Coverage_by_metric: CI coverage rate (%) by main strategy and population-generating characteristics
Given the strategies' approximate unbiasedness and near-nominal CI coverage, we focus on comparing strategies in terms of RMSE, which we generally consider relative to the B-P strategy.
Increases in relative RMSE were paralleled by very similar proportional increases in CI relative width; therefore, we do not report the latter.

Tables \ref{tab:RelIncrease_RMSE_NeymanHT} and \ref{tab:RelIncrease_RMSE_CochranSR} display the relative increase in RMSE of N-HT and C-SR, respectively, compared with B-P, for each of the 90 populations.
Rows reflect different levels of heteroscedasticity used in generating the populations, panels reflect the three size measures used, and columns within a panel reflect the six scenarios for strata correlations and slopes.

%For subsequent versions, consider moving Tables 2-3 back here
%I manually moved them down a few paragraphs since otherwise the 
% three paragraphs are unnecessarily displayed on their own page.

Across all populations, N-HT consistently underperformed B-P, increasing the RMSE by 11\%--175\% for individual populations (Table \ref{tab:RelIncrease_RMSE_NeymanHT}).
The increases in relative RMSE were greatest for the \textit{higher correlations} scenarios (122\%--175\%) and tended to be smallest for the \textit{lower correlations} scenarios (11\%--44\%).
Similar patterns were observed when comparing N-HT with C-SR, although the magnitudes were sometimes smaller.
The advantages of C-SR and B-P over N-HT were not surprising, since the latter does not incorporate the auxiliary variable in estimation.

Further, B-P performed about as well or better than C-SR, but with marked differences across populations, with B-P showing the greatest advantage for a subset of MOS1 scenarios, wherein the true value of $b$ was equal to 2 or 0 (Table \ref{tab:RelIncrease_RMSE_CochranSR}).
Specifically, C-SR resulted in relative increases in RMSE of 21\%--29\% for the MOS1, $b=2$ populations, and 10\%--40\% for the MOS1, $b=0$ populations.
In contrast, C-SR only yielded a 3.1\% average increase in RMSE across the MOS2 populations and 2.0\% for the MOS3 populations, which were not as heavily skewed.

For the populations where $b\ge 0.5$, B-P tended to produce more stable sample allocations than C-SR with respect to repeated pilot sampling, and increasingly so for higher levels of $b$ and/or increased skewness of X.
The differences were starkest for the MOS1, $b=2$ populations, as displayed in Table \ref{table:paper1:sim_allocation_summary_statistics}, where the CVs for the B-P allocations were appreciably lower than those of the alternatives.
Note that the B-P allocation incorporates more detailed population information than does C-SR, the latter of which is solely based on estimated population residuals and does not directly incorporate $b$.

\begin{footnotesize}
\captionsetup{font=small}

\begin{landscape}
	\begin{singlespace}
		\newpage
		
\begin{longtable}[l]{>{}l|rrrrr>{}r|rrrrr>{}r|rrrrr>{}r|}
\caption{\label{tab:RelIncrease_RMSE_NeymanHT}Relative increase in RMSE from Neyman-HT versus B-P (\%)}\\
\toprule
\multicolumn{1}{c}{ } & \multicolumn{6}{c}{MOS1 (most skewness)} & \multicolumn{6}{c}{MOS2 (mid skewness)} & \multicolumn{6}{c}{MOS3 (least skewness)} \\
\cmidrule(l{3pt}r{3pt}){2-7} \cmidrule(l{3pt}r{3pt}){8-13} \cmidrule(l{3pt}r{3pt}){14-19}
\rotatebox{90}{} & \rotatebox{90}{1. Baseline} & \rotatebox{90}{2. Lower corrs} & \rotatebox{90}{3. Higher corrs} & \rotatebox{90}{4. Inc. corrs} & \rotatebox{90}{5. Dec. slopes} & \rotatebox{90}{6. Inc. corrs, dec. slopes} & \rotatebox{90}{1. Baseline} & \rotatebox{90}{2. Lower corrs} & \rotatebox{90}{3. Higher corrs} & \rotatebox{90}{4. Inc. corrs} & \rotatebox{90}{5. Dec. slopes} & \rotatebox{90}{6. Inc. corrs, dec. slopes} & \rotatebox{90}{1. Baseline} & \rotatebox{90}{2. Lower corrs} & \rotatebox{90}{3. Higher corrs} & \rotatebox{90}{4. Inc. corrs} & \rotatebox{90}{5. Dec. slopes} & \rotatebox{90}{6. Inc. corrs, dec. slopes}\\
\midrule
b = 0 & 65.4 & 31.8 & 175.3 & 72.2 & 69.4 & 77.3 & 37.7 & 24.9 & 142.1 & 45.6 & 54.3 & 37.8 & 43.6 & 18.4 & 133.0 & 39.3 & 41.4 & 41.3\\
b = 0.5 & 44.3 & 18.4 & 138.4 & 43.7 & 50.8 & 36.1 & 40.8 & 21.8 & 133.8 & 41.4 & 42.1 & 36.9 & 39.7 & 21.1 & 129.9 & 32.0 & 30.7 & 29.6\\
b = 1 & 46.0 & 24.5 & 127.4 & 34.9 & 40.5 & 27.8 & 45.0 & 14.1 & 135.2 & 36.7 & 42.3 & 32.0 & 35.0 & 18.9 & 126.2 & 34.2 & 34.2 & 36.0\\
b = 1.5 & 51.9 & 24.4 & 139.5 & 41.0 & 50.9 & 38.2 & 37.2 & 11.5 & 131.5 & 41.1 & 45.1 & 31.3 & 40.0 & 21.5 & 134.2 & 41.1 & 41.3 & 27.9\\
b = 2 & 63.9 & 43.7 & 158.7 & 74.6 & 67.9 & 61.7 & 47.4 & 13.0 & 121.9 & 40.0 & 46.5 & 40.5 & 42.0 & 19.2 & 133.1 & 41.0 & 39.2 & 41.6\\
\bottomrule
\end{longtable}

\begin{longtable}[l]{>{}l|rrrrr>{}r|rrrrr>{}r|rrrrr>{}r|}
\caption{\label{tab:RelIncrease_RMSE_CochranSR}Relative increase in RMSE from Cochran-SR versus B-P (\%)}\\
\toprule
\multicolumn{1}{c}{ } & \multicolumn{6}{c}{MOS1 (most skewness)} & \multicolumn{6}{c}{MOS2 (mid skewness)} & \multicolumn{6}{c}{MOS3 (least skewness)} \\
\cmidrule(l{3pt}r{3pt}){2-7} \cmidrule(l{3pt}r{3pt}){8-13} \cmidrule(l{3pt}r{3pt}){14-19}
\rotatebox{90}{} & \rotatebox{90}{1. Baseline} & \rotatebox{90}{2. Lower corrs} & \rotatebox{90}{3. Higher corrs} & \rotatebox{90}{4. Inc. corrs} & \rotatebox{90}{5. Dec. slopes} & \rotatebox{90}{6. Inc. corrs, dec. slopes} & \rotatebox{90}{1. Baseline} & \rotatebox{90}{2. Lower corrs} & \rotatebox{90}{3. Higher corrs} & \rotatebox{90}{4. Inc. corrs} & \rotatebox{90}{5. Dec. slopes} & \rotatebox{90}{6. Inc. corrs, dec. slopes} & \rotatebox{90}{1. Baseline} & \rotatebox{90}{2. Lower corrs} & \rotatebox{90}{3. Higher corrs} & \rotatebox{90}{4. Inc. corrs} & \rotatebox{90}{5. Dec. slopes} & \rotatebox{90}{6. Inc. corrs, dec. slopes}\\
\midrule
b = 0 & 19.4 & 9.7 & 18.1 & 32.4 & 28.5 & 40.5 & -0.8 & 7.1 & 5.6 & 13.5 & 7.3 & 2.6 & 2.2 & 0.8 & -0.8 & 1.6 & 4.8 & 4.3\\
b = 0.5 & 7.1 & 3.0 & 0.2 & 2.1 & 11.6 & 7.5 & 1.6 & -2.4 & 8.4 & 2.6 & 1.2 & 3.6 & -0.5 & 7.5 & 2.4 & -0.2 & 2.7 & -1.7\\
b = 1 & 5.8 & 7.4 & 3.2 & 1.0 & 4.3 & 3.4 & -1.3 & -1.2 & 1.4 & -2.3 & -0.3 & 7.2 & -3.4 & 3.6 & -0.2 & 1.1 & -3.6 & 3.9\\
b = 1.5 & 11.0 & 9.1 & 5.1 & 5.3 & 5.9 & 5.3 & 3.6 & 0.8 & 2.9 & 5.2 & 0.5 & 6.0 & 2.3 & -1.7 & 2.4 & 3.9 & 2.3 & 0.5\\
b = 2 & 22.7 & 21.1 & 23.0 & 24.6 & 24.2 & 28.5 & 3.1 & 4.2 & 1.6 & 1.2 & 4.0 & 6.9 & 1.8 & 4.6 & 5.8 & 2.6 & 2.0 & 7.9\\
\bottomrule
\end{longtable}

	\end{singlespace}
\end{landscape}
\end{footnotesize}

\newpage	\begin{singlespace}

\begin{longtable}[l]{l>{}r|r>{}r|r>{}r|rr}
\caption{\label{tab:unnamed-chunk-24}Allocation summary statistics by scenario, stratum, and strategy: populations 25 to 30 (MOS1, b=2)}\\
\toprule
\multicolumn{2}{c}{ } & \multicolumn{2}{c}{Neyman-HT} & \multicolumn{2}{c}{Cochran-SR} & \multicolumn{2}{c}{Bayesian-P} \\
\cmidrule(l{3pt}r{3pt}){3-4} \cmidrule(l{3pt}r{3pt}){5-6} \cmidrule(l{3pt}r{3pt}){7-8}
Scenario & h & mean & sd & mean & sd & mean & sd\\
\midrule
 & 1 & 148.7 & 47.9 & 134.9 & 50.4 & 111.6 & 18.8\\
\nopagebreak
 & 2 & 90.4 & 19.9 & 92.2 & 21.8 & 98.1 & 16.7\\
\nopagebreak
 & 3 & 75.5 & 16.4 & 80.3 & 18.8 & 85.2 & 15.1\\
\nopagebreak
 & 4 & 84.5 & 16.8 & 85.1 & 18.9 & 91.2 & 15.3\\
\nopagebreak
\multirow{-5}{*}{\raggedright\arraybackslash 1. Baseline} & 5 & 100.9 & 21.6 & 107.4 & 24.2 & 113.9 & 19.1\\
\cmidrule{1-8}\pagebreak[0]
 & 1 & 147.1 & 50.4 & 135.5 & 52.0 & 114.0 & 18.5\\
\nopagebreak
 & 2 & 91.8 & 21.6 & 94.8 & 22.4 & 97.9 & 16.2\\
\nopagebreak
 & 3 & 75.7 & 18.0 & 77.7 & 18.3 & 83.3 & 14.5\\
\nopagebreak
 & 4 & 82.8 & 18.2 & 84.2 & 19.3 & 89.7 & 15.3\\
\nopagebreak
\multirow{-5}{*}{\raggedright\arraybackslash 2. Lower corrs} & 5 & 102.7 & 22.3 & 107.8 & 23.7 & 115.1 & 18.4\\
\cmidrule{1-8}\pagebreak[0]
 & 1 & 154.3 & 40.4 & 133.4 & 51.1 & 112.2 & 19.1\\
\nopagebreak
 & 2 & 87.5 & 16.9 & 93.3 & 22.3 & 96.8 & 15.7\\
\nopagebreak
 & 3 & 74.5 & 13.8 & 78.6 & 17.8 & 83.5 & 14.2\\
\nopagebreak
 & 4 & 84.0 & 15.4 & 90.6 & 20.8 & 96.5 & 15.6\\
\nopagebreak
\multirow{-5}{*}{\raggedright\arraybackslash 3. Higher corrs} & 5 & 99.7 & 17.2 & 104.1 & 21.1 & 110.9 & 16.5\\
\cmidrule{1-8}\pagebreak[0]
 & 1 & 189.6 & 56.5 & 206.8 & 64.7 & 183.2 & 25.2\\
\nopagebreak
 & 2 & 96.5 & 24.2 & 111.9 & 31.0 & 118.9 & 20.0\\
\nopagebreak
 & 3 & 72.8 & 18.4 & 77.7 & 22.7 & 84.4 & 15.6\\
\nopagebreak
 & 4 & 67.7 & 16.6 & 59.4 & 17.5 & 64.8 & 11.9\\
\nopagebreak
\multirow{-5}{*}{\raggedright\arraybackslash 4. Inc. corrs} & 5 & 73.4 & 17.3 & 44.2 & 13.0 & 48.7 & 9.2\\
\cmidrule{1-8}\pagebreak[0]
 & 1 & 193.7 & 53.1 & 174.8 & 59.9 & 154.7 & 23.0\\
\nopagebreak
 & 2 & 103.8 & 24.6 & 108.4 & 28.7 & 113.8 & 19.7\\
\nopagebreak
 & 3 & 79.4 & 18.4 & 82.1 & 20.3 & 87.2 & 14.3\\
\nopagebreak
 & 4 & 63.4 & 15.4 & 68.5 & 17.5 & 73.6 & 13.2\\
\nopagebreak
\multirow{-5}{*}{\raggedright\arraybackslash 5. Dec. slopes} & 5 & 59.8 & 14.4 & 66.2 & 17.1 & 70.7 & 12.6\\
\cmidrule{1-8}\pagebreak[0]
 & 1 & 232.5 & 60.2 & 242.9 & 66.7 & 220.7 & 26.4\\
\nopagebreak
 & 2 & 108.8 & 29.5 & 122.0 & 37.6 & 129.8 & 21.9\\
\nopagebreak
 & 3 & 65.2 & 18.7 & 67.3 & 22.0 & 73.2 & 14.0\\
\nopagebreak
 & 4 & 51.5 & 14.9 & 42.8 & 14.1 & 47.6 & 9.5\\
\nopagebreak
\multirow{-5}{*}{\raggedright\arraybackslash 6. Inc. corrs, dec. slopes} & 5 & 41.9 & 11.5 & 25.0 & 8.5 & 28.8 & 5.7\\
\bottomrule
\end{longtable}

{\makeatletter\def\@currentlabel{\thetable}\label{table:paper1:sim_allocation_summary_statistics}}
\end{singlespace}

\subsection{Rule-of-thumb allocations}

Next, we examined performance of rule-of-thumb strategies in populations where they are potentially appropriate (i.e., RT0-SR, RT1-SR, and RT2-SR for the $b=0$, $b=1$, and $b=2$ populations, respectively).
The C-SR and B-P strategies, which incorporate pilot data when computing allocations, consistently performed as well or better than the RT-SR strategies, which do not.
For example, for the MOS1, $b=2$ populations, RT2-SR led to 82\%--204\% increases in RMSE compared with B-P (Table \ref{tab:RelIncrease_RMSE_vs_BP}, first panel, third row), with the worst performance for the scenario with decreasing slopes and increasing correlations.
In contrast, RT1-SR performed similarly or only modestly worse than the other strategies in the nine $b=1$ populations with constant correlations and slopes (i.e., MOS1, MOS2, and MOS3 populations under baseline, lower, or higher correlation scenarios).

\begin{footnotesize}
\begin{singlespace}
	
\begin{longtable}[l]{>{}l|rrrrr>{}r|rrrrr>{}r|rrrrr>{}r|}
\caption{\label{tab:RelIncrease_RMSE_vs_BP}Relative increase in RMSE from RT-SR strategies versus B-P (\%)}\\
\toprule
\multicolumn{1}{c}{ } & \multicolumn{6}{c}{MOS1 (most skewness)} & \multicolumn{6}{c}{MOS2 (mid skewness)} & \multicolumn{6}{c}{MOS3 (least skewness)} \\
\cmidrule(l{3pt}r{3pt}){2-7} \cmidrule(l{3pt}r{3pt}){8-13} \cmidrule(l{3pt}r{3pt}){14-19}
\rotatebox{90}{} & \rotatebox{90}{1. Baseline} & \rotatebox{90}{2. Lower corrs} & \rotatebox{90}{3. Higher corrs} & \rotatebox{90}{4. Inc. corrs} & \rotatebox{90}{5. Dec. slopes} & \rotatebox{90}{6. Inc. corrs, dec. slopes} & \rotatebox{90}{1. Baseline} & \rotatebox{90}{2. Lower corrs} & \rotatebox{90}{3. Higher corrs} & \rotatebox{90}{4. Inc. corrs} & \rotatebox{90}{5. Dec. slopes} & \rotatebox{90}{6. Inc. corrs, dec. slopes} & \rotatebox{90}{1. Baseline} & \rotatebox{90}{2. Lower corrs} & \rotatebox{90}{3. Higher corrs} & \rotatebox{90}{4. Inc. corrs} & \rotatebox{90}{5. Dec. slopes} & \rotatebox{90}{6. Inc. corrs, dec. slopes}\\
\midrule
b = 0 & 45 & 34 & 43 & 35 & 38 & 29 & 4 & 10 & 11 & 12 & 7 & 11 & 6 & 10 & 0 & 3 & -1 & 12\\
b = 1 & 5 & 10 & 4 & 18 & 10 & 32 & 3 & 3 & 7 & 17 & 11 & 37 & -4 & 1 & 5 & 18 & 9 & 40\\
b = 2 & 91 & 84 & 82 & 148 & 133 & 204 & 25 & 31 & 23 & 72 & 55 & 105 & 21 & 17 & 23 & 55 & 39 & 85\\
\bottomrule
\end{longtable}

\end{singlespace}
\end{footnotesize}

Otherwise, the RT-SR strategies typically had little bias and near-nominal coverage of 95\% CIs (excepting the MOS1, $b=2$ populations, where coverage was closer to 90\%; Appendix C.2, Table 4).
%\ref{tab:cochran_RT_relbias_coverage}).
%tab:cochran_RT_relbias_coverage
%RT-SR results: 1,000*RelBias and CI coverage (%), by population

\subsection{Effects of misspecified b on B-P strategy}
\label{subsec:ch2:simulation_misspecified_b}

\begin{figure}[H]
\centering
\caption{Average sample allocations by population and assumed b, selected populations (MOS1, baseline $\{\alpha_h,\rho_h\}$ scenario)}
\label{fig:misspec_b_avg_alloc_by_pop}
\includegraphics[width=0.7\linewidth]{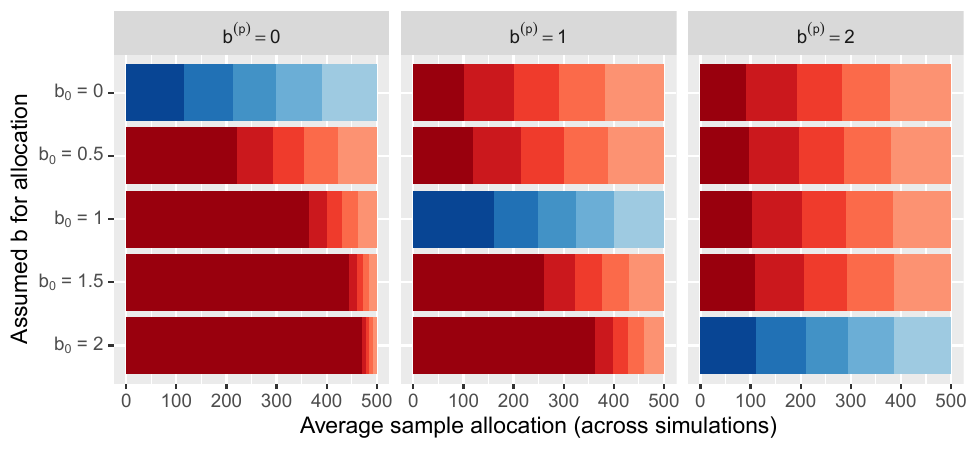}
\includegraphics[width=0.7\linewidth]{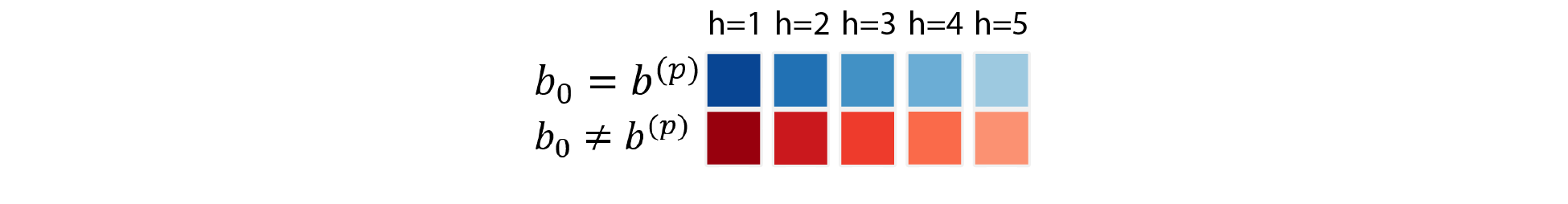}
\end{figure}

Next, we evaluated the effects of misspecifying $b$ at the sample allocation stage when using the B-P strategy---that is, allocation via $b_0 \in \{0, 0.5, 1, 1.5, 2\}$, but where $b_0 \ne b^{(p)}$.
The results varied widely across populations, in terms of RMSE and average allocations.
In the MOS2 and MOS3 populations, the B-P strategy was fairly insensitive to misspecified $b$; in contrast, some MOS1 populations were very sensitive to this, especially at the lowest levels of $b$.
To illustrate, Figure \ref{fig:misspec_b_avg_alloc_by_pop} shows the average allocations to sampling strata (via Equation \ref{eqn:sim_design_avg_alloc}) for three MOS1 populations, under the baseline $\{\alpha_h,\rho_h\}$ scenario, where $b^{(p)}$ equaled 0, 1, or 2, and where correctly specified allocations are colored in blue (with shade denoting stratum, and bar width denoting average allocation size).
For the $b^{(p)}=0$ population (leftmost panel), the correct specification ($b_0=0$; top row, in blue) led to a roughly equal allocation, whereas overestimates of $b_0$ led to substantial overallocations for stratum 1 (e.g., leftmost panel, bottom row, in red).
In contrast, for the corresponding $b^{(p)}=2$ population (rightmost panel), underestimates of $b$ led to only slight changes in allocation, on average.

As a result, among MOS1 populations, the RMSEs of the B-P strategy were very susceptible to misspecification of $b$ when $b^{(p)}=0$, with this effect diminishing at higher levels of $b^{(p)}$.
This is illustrated in Table \ref{tab:RelIncrease_RMSE_misspec_b}, which provides the relative increase in RMSE from misspecified $b$ for selected MOS1 populations.

Beyond the main results above, misspecification under these MOS1 populations still allowed for approximately unbiased estimates, with 95\% CIs that usually had reasonable coverage (albeit with overly conservative CIs when $b^{(p)}=0$ and $b_0 \in \{1.5,2\}$).

\begin{landscape}
\begin{singlespace}
	\begin{table}

\caption{\label{tab:RelIncrease_RMSE_misspec_b}Relative increase in RMSE from misspecified b versus correctly specified b among selected MOS1 populations (\%)}
\resizebox{\linewidth}{!}{
\begin{tabular}[t]{>{}l|rrrrr>{}r|rrrrr>{}r|rrrrr>{}r|}
\toprule
\multicolumn{1}{c}{ } & \multicolumn{6}{c}{$b^{(p)} = 0$} & \multicolumn{6}{c}{$b^{(p)} = 1$} & \multicolumn{6}{c}{$b^{(p)} = 2$} \\
\cmidrule(l{3pt}r{3pt}){2-7} \cmidrule(l{3pt}r{3pt}){8-13} \cmidrule(l{3pt}r{3pt}){14-19}
\rotatebox{90}{ } & \rotatebox{90}{1. Baseline} & \rotatebox{90}{2. Lower corrs} & \rotatebox{90}{3. Higher corrs} & \rotatebox{90}{4. Inc. corrs} & \rotatebox{90}{5. Dec. slopes} & \rotatebox{90}{6. Inc. corrs, dec. slopes} & \rotatebox{90}{1. Baseline} & \rotatebox{90}{2. Lower corrs} & \rotatebox{90}{3. Higher corrs} & \rotatebox{90}{4. Inc. corrs} & \rotatebox{90}{5. Dec. slopes} & \rotatebox{90}{6. Inc. corrs, dec. slopes} & \rotatebox{90}{1. Baseline} & \rotatebox{90}{2. Lower corrs} & \rotatebox{90}{3. Higher corrs} & \rotatebox{90}{4. Inc. corrs} & \rotatebox{90}{5. Dec. slopes} & \rotatebox{90}{6. Inc. corrs, dec. slopes}\\
\midrule
$b_0 = 0$ &  &  &  &  &  &  & 7 & 16 & 10 & 12 & 5 & 13 & 3 & 7 & 3 & 5 & 8 & 3\\
$b_0 = 0.5$ & 11 & 2 & 16 & 17 & 13 & 20 & 7 & 8 & 1 & 6 & 5 & 5 & 5 & 7 & 2 & 5 & 2 & -2\\
$b_0 = 1$ & 72 & 65 & 87 & 76 & 75 & 86 &  &  &  &  &  &  & 4 & 2 & 0 & 2 & 3 & -1\\
$b_0 = 1.5$ & 182 & 177 & 204 & 193 & 200 & 187 & 8 & 13 & 7 & 8 & 8 & 10 & 3 & -1 & -5 & 1 & -3 & -3\\
$b_0 = 2$ & 271 & 267 & 281 & 265 & 268 & 240 & 53 & 56 & 48 & 46 & 53 & 46 &  &  &  &  &  & \\
\bottomrule
\end{tabular}}
\end{table}

	{\makeatletter\def\@currentlabel{\thetable}\label{table:RelIncrease_RMSE_misspec_b}}
\end{singlespace}
\end{landscape}

\section{Application using NCCS data}
\label{sec:ch2:NCCS_simulation}

\subsection{Analysis of log revenues}
\label{sec:ch2:NCCS_simulation_logrev} 
Next, we applied our methods to analyzing tax returns of public charities, using data from the National Center for Charitable Statistics (NCCS), and which were largely based on IRS Form 990 data.
Using NCCS Core 1989--2015 Public Charities Fiscal Year Trend data, we analyzed total revenues in fiscal years 2008 and 2013, using Employer Identification Number (EIN) as an identifier, and focusing on organizations with at least \$100,000 in 2008 revenue.
We focused on the U.S. nonprofit sector, that is, excluding records that NCCS flags as out-of-scope (e.g., due to being overseas or in U.S. territories).
We focused on operating public charities (i.e., as opposed to supporting or mutual benefit public charities), to avoid double counting.
Further, we focused on organizations with at least \$50,000 in gross receipts in each year, to reflect filing thresholds (which were raised to that level in 2010).
We also required positive revenues, so that we could subsequently consider log revenue.
This resulted in a population of 140,858 domestic operating public charities with at least \$100,000 in 2008 revenue, and positive revenue in 2013, and who met filing thresholds.
We subsequently categorized cases by nonprofit sector, based on National Taxonomy of Exempt Entities (NTEE) codes; therefore, we dropped an additional 65 cases with missing NTEE category.

\begin{figure}[h]
\centering
\caption{Contour plot for kernel density estimates of log revenue in 2008 versus 2013, by sector}
\includegraphics[width=0.7\linewidth]{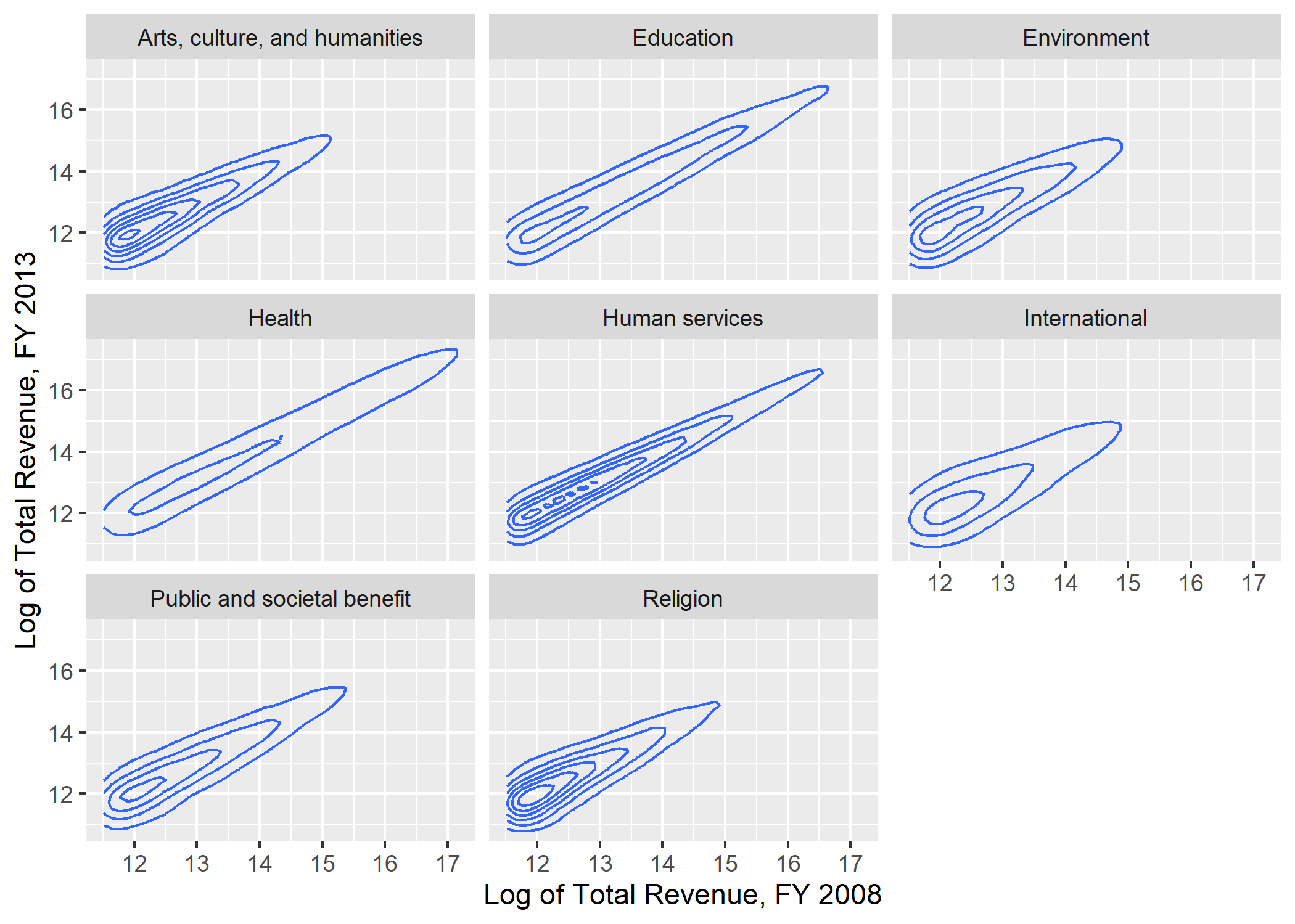}
\label{fig:logrevbysector2ddensityplot}
\end{figure}

Using the resulting data, we aimed to estimate the total log revenue in fiscal year 2013, using log revenue in 2008 as an auxiliary variable.
Figure \ref{fig:logrevbysector2ddensityplot} displays the contours of 2D kernel density estimates of the data, by sector.

Next, we estimated $b$ via MCMC in \textsf{JAGS} \citep{Plummer2017}, via \textsf{R}.
In doing so, we assumed an unstratified model of $Y_i = \alpha X_i + \varepsilon_i$, where $\varepsilon_i \overset{ind}\sim N(0, \tau / X_i^b)$ (the second parameter being variance), using a normal prior on $\alpha$, inverse-gamma prior on $\tau$, and uniform prior on $b$ (with support from -3 to 3).
After 2,000 burn-in iterations and simulating an additional 20,000 iterations for each of five chains, we retained every tenth observation for a total of 10,000 retained simulations.
MCMC diagnostics were plotted using \texttt{ggmcmc} \citep{Fernandez2016ggmcmc}, including the potential scale reduction factor \citep[$\hat{R}$,][]{Gelman2003}, which suggested approximate convergence.
This analysis yielded a posterior mean of $\hat{b}=0.55$ and 95\% equal-tail CI for $b$ of $(0.25, 0.66)$

Next, we used methods that paralleled those of our earlier simulation, as applied to the NCCS data.
We formed 24 strata based on the cross-classification of nonprofit sector (8 categories) and 2008 revenue (3 categories), using revenue boundaries that roughly equalized the aggregate unit standard deviations for the three size categories, as applied to log revenue in 2008, and assuming that $b= 0.55$.
Population counts for the resulting strata are displayed in Table \ref{table:irs_popsizes_sim}.
We then drew $R=10,000$ equally allocated pilot samples of $n=360$ units (15 per stratum), each of which was used in applying the strategies of Table \ref{table:overview_of_sim_strategies} for drawing main studies with total sample sizes of $n=1800$.
Strategies were again compared in terms of RMSE, relative bias, and CI properties (see $\S\ref{subsec:Simulation:performance_comparison}$).

\begin{singlespace}

\begin{longtable}[t]{>{}l|rr>{}r|r}
\caption{\label{tab:unnamed-chunk-10}NCCS data: population counts by sector and 2008 revenue}\\
\toprule
\multicolumn{1}{c}{ } & \multicolumn{3}{c}{Total Revenue in 2008} & \multicolumn{1}{c}{ } \\
\cmidrule(l{3pt}r{3pt}){2-4}
Nonprofit Sector & Under \$300k & \$300k--\$1.2m & \$1.2m+ & Total\\
\midrule
\endfirsthead
\caption[]{NCCS data: population counts by sector and 2008 revenue \textit{(continued)}}\\
\toprule
\multicolumn{1}{c}{ } & \multicolumn{3}{c}{Total Revenue in 2008} & \multicolumn{1}{c}{ } \\
\cmidrule(l{3pt}r{3pt}){2-4}
Nonprofit Sector & Under \$300k & \$300k--\$1.2m & \$1.2m+ & Total\\
\midrule
\endhead

\endfoot
\bottomrule
\endlastfoot
Arts, culture, and humanities & 6,533 & 4,883 & 2,870 & 14,286\\
Education & 5,652 & 5,897 & 7,882 & 19,431\\
Environment & 2,633 & 2,219 & 1,308 & 6,160\\
Health & 4,542 & 5,668 & 11,706 & 21,916\\
Human services & 19,929 & 19,896 & 17,795 & 57,620\\
International & 1,179 & 959 & 904 & 3,042\\
Public and societal benefit & 4,125 & 3,862 & 2,876 & 10,863\\
Religion & 3,822 & 2,498 & 1,155 & 7,475\\
\midrule
Total & 48,415 & 45,882 & 46,496 & 140,793\\*
\end{longtable}

{\makeatletter\def\@currentlabel{\thetable}\label{table:irs_popsizes_sim}}
\end{singlespace}

Simulation results are summarized in Table \ref{table:Results_IRS990}.
The C-SR and B-P strategies offered substantially lower RMSE than the N-HT strategy.
All three methods were approximately unbiased and had near-nominal CI coverage, the B-P strategy producing slightly conservative CIs.

\begin{singlespace}
\begin{table}[!h]

\caption{NCCS simulation results}
\centering
\begin{tabular}[t]{>{}l|cccc}
\toprule
Strategy & Relative RMSE & 1000*RelBias & CI Coverage (\%) & 1000*CI RelWidth\\
\midrule
N-HT & 1.427 & -0.00 & 94.7 & 6.48\\
C-SR & 1.014 & -0.01 & 94.8 & 4.57\\
B-P & 1.000 & -0.00 & 95.5 & 4.63\\
\bottomrule
\multicolumn{5}{l}{\rule{0pt}{1em}\textit{Note: } RMSE is displayed relative to that of the B-P strategy.}\\
\end{tabular}
\end{table}

{\makeatletter\def\@currentlabel{\thetable}\label{table:Results_IRS990}}
\end{singlespace}

\subsection{Analysis on original scale}
\label{sec:ch2:NCCS_simulation_origscale}
In practice, samples are sometimes allocated under models that do not exactly fit the data.
Therefore, an important question concerns the performance of B-P methods for scenarios where the model does not fit well.
We analyzed such a scenario by repeating the simulation of \S\ref{sec:ch2:NCCS_simulation_logrev}, but using the NCCS data on their original scale (i.e., omitting the log transformation), which resulted in a poor fit for our assumed model.

Among the 140,793 charities analyzed in \S\ref{sec:ch2:NCCS_simulation_logrev}, 1.3\% of units had 2008 revenues exceeding \$100 million, which totaled 64\% of total 2008 revenues.
Considering that our focus is not on extremely large units, which would presumably be selected under any reasonable design, we omitted these units from our analysis, and analyzed the remaining $N = 138,960$ charities.
We assumed that the inferential objective is to estimate total revenue in fiscal year 2013, using 2008 revenue as the auxiliary variable.

We again estimated $b$ via MCMC, as in \S\ref{sec:ch2:NCCS_simulation_logrev}, but where the $X$ and $Y$ reflect the original scale, which resulted in a posterior mean of $\hat{b} = 1.19$, which we assumed to be the true value of $b$.
%Posterior mean = 1.194176; 
%Quantiles:
%    2.5%      50%    97.5% 
%1.189361 1.194114 1.199514 
Even accounting for heteroscedasticity, we note that the normal distribution is a poor approximation for the underlying data.
This is evident from a Q-Q plot for a simple (no-intercept) linear regression on 2013 revenue on 2008 revenue, with analytical weights of $1/X_i^{1.19}$ (see Appendix C.2, Figure 1, %\ref{fig:app_irs_qqplot_faceted}
righthand panel), which has a far more extreme righthand tail than an analogous plot for the log-transformed data (lefthand panel).
%	\caption{Q-Q plots for unstratified regressions of IRS data, based on log-transformed and original scales}
%{fig:app_irs_qqplot_faceted} 

We formed 24 strata in the same manner as before, except that the size class boundaries were modified to reflect the current section's estimate of $b$ (see Appendix C.2, Table 6 %\ref{table:irs_popsizes_sim_100m} 
%table:irs_popsizes_sim_100m -- NCCS data: population counts by sector and 2008 revenue (with size classes based on original scale)
for population counts that reflect the current boundaries).
Given that some of the resulting strata population sizes were small, implementing the classic allocations directly would have led to some allocations of $n_h < 4$ or $n_h > N_h$.
Therefore, allocations below 4 or above $N_h$ were moved to these bounds; this step was applied in conjunction with the rounding step described in the last paragraph of \S\ref{subsec:Simulation:mainstudy}, so that the total sample size was always $n = 1800$.

Simulation results are shown in Table \ref{table:Results_IRS990_origscale}.
The C-SR and B-P strategies again outperformed N-HT, but more narrowly.
All strategies produced CIs with slightly below nominal coverage.

\begin{singlespace}
\begin{table}[!h]

\caption{NCCS simulation results: original scale}
\centering
\begin{tabular}[t]{>{}l|cccc}
\toprule
Strategy & Relative RMSE & 1000*RelBias & CI Coverage (\%) & 1000*CI RelWidth\\
\midrule
N-HT & 1.123 & 0.27 & 93.2 & 11.46\\
C-SR & 0.994 & -0.24 & 90.2 & 9.55\\
B-P & 1.000 & 2.48 & 91.7 & 9.65\\
\bottomrule
\multicolumn{5}{l}{\rule{0pt}{1em}\textit{Note: } RMSE is displayed relative to that of the B-P strategy.}\\
\end{tabular}
\end{table}

{\makeatletter\def\@currentlabel{\thetable}\label{table:Results_IRS990_origscale}}
\end{singlespace}

\section{Discussion}
\label{sec:ch2:discussion}

\subsection{Summary of results}

We considered sample allocation for stratified random sampling under a univariate regression model with heteroscedastic errors, using Bayesian decision theory to accommodate uncertain design parameters.
We focused on optimal allocation for estimating the finite population mean, under squared error loss.
We assumed a non-informative prior in conjunction with use of a pilot survey, so that the allocation would be driven by the pilot data rather than via a subjective prior.
We derived the approximate expected posterior variance, which allowed for identifying the optimal allocation via mathematical programming.
This entailed selecting the $\{n_h\}$ to minimize the approximate expected posterior variance, subject to constraints of $4 \le n_h \le N_h$, although additional constraints on sample sizes or other sample-related quantities could have been straightforwardly handled via standard optimization software \citep[see, e.g.,][chap. 5]{Valliant2013}.
Although the expected posterior loss assumed large $\{n_h\}$ in applying a Taylor series approximation, the resulting empirical performance was nearly identical to that of an alternative that avoided this assumption.

Performance was compared with key design-based and model-assisted strategies across a series of artificial populations with a skewed, continuous size measure, and outcome variables that followed from various applications of our model.
Among the main strategies considered, the model-assisted method (C-SR) and proposed Bayesian method (B-P), which incorporate auxiliary information in estimation, consistently outperformed the purely design-based strategy (N-HT), which did not.
B-P performed as well or better than C-SR across the different populations, achieving similar or moderately lower RMSEs; for certain types of populations, B-P also led to more stable sample allocations with respect to repeated pilot sampling.

Although the simulations above assumed use of our model, we repeated the simulations using real data.
In an application to the log-revenues of public charities, the B-P strategy achieved substantially reduced RMSE compared with N-HT, as did the C-SR strategy.
Likewise, B-P and C-SR turned out to outperform N-HT when the public charities application was repeated using the original scale, in spite of a worsened model fit.

We considered rule-of-thumb (RT) strategies in populations where they were potentially appropriate.
The C-SR and B-P strategies, which incorporate pilot data, consistently performed as well as or better than the RT strategies, whose allocations are solely based on the auxiliary variable; in some cases, the RT strategies performed substantially worse, particularly for $b=2$ populations.

We also evaluated the effects of misspecified coefficient of heteroscedasticity at the sample allocation stage under the proposed strategy.
We found that under a highly skewed size measure, and when the true $b$ was close to 0, the sample allocations were very sensitive to misspecified $b$.
On the other hand, for many populations considered, misspecification of $b$ did not meaningfully change the resulting allocations or RMSE.

\subsection{Limitations and future directions}
\label{sec:ch2:limitations_future_directions}

Although our allowing for heteroscedasticity improves realism in comparison to some previous Bayesian optimal design work, our model assumptions will not always be realistic.
Therefore, an important avenue of future work entails development of alternative models, such as
multivariate models, different error terms (e.g., lognormal), and/or different conditional mean functions.
Some examples of the latter are from the class of general polynomial models considered by \citet[][53--54]{Valliant2000} and from the set of associations considered in simulation by \citet{Zangeneh2015}.
It may also be important to consider the implications of outliers or other influential values on allocations.
Of course, changes to the population structure may require new derivations for the posterior expected loss, or alternatively, the development of a computational approach \citep[e.g.,][]{Ryan2016}.

An important aspect of the problem structure is that separate superpopulation parameters are used for different strata.
If there are too many strata relative to the total sample size, then the resulting estimator will be inefficient, since the stratum-specific predictions will be unstable.
As an extreme example, considering the $(n_h-1)/(n_h-3)$ term in (\ref{eqn:EV_fixed_b_final}), and assuming a target sample size of $n$, using $H = n/4$ strata would result in four units per stratum regardless of how the strata were formed.
Even if the strata are large enough such that most of the population is in strata where $(n_h-1)/(n_h-3)\approx 1$, if there are sets of strata with similar model coefficients, then it may be more efficient to pool groups of strata.
On the other extreme, having far too few strata may lead to the use of a model that does not adequately capture group differences.
Further work could be done to explore these issues.

Although the B-P approach performed as well or better than the alternatives, most of the populations used in simulation were generated according to our model, as was necessary due to the paucity of publicly available establishment data.
When repeating the simulation to public charities data, the B-P strategy performed well, but so did the C-SR strategy, which had roughly similar RMSE.
The simulations on artificial data show that the B-P strategy can sometimes improve on C-SR under a correctly specified model, whereas the application to public charities data showed the B-P strategy performing reasonably in an application to real data.
Additional empirical or theoretical work could be conducted to better understand the conditions under which the methods' performance varies and the reasons why.

Additionally, given that design-based inference is predominant at most statistical agencies---notwithstanding an increasing reliance on models in recent years to combat nonsampling errors \citep[e.g.][]{Valliant2022_MH_jssam}---the reader may wonder about how our proposed optimal Bayes methods relate to design-based inference.
While the proposed Bayes estimator relies on Bayesian methodology to develop the sample design, the Bayesian piece---the prior distribution---is based on a simple non-informative prior that often can be viewed as an inferential equivalent to frequentist estimators.  Further, the evaluations in this manuscript follow the “calibrated Bayes” paradigm \citep{Little2006} of using Bayesian methods to develop estimators but to evaluate these estimators using frequentist methods (bias, mean square error).

We assumed that $b$ is known upfront, which facilitated analytical results, but this assumption may sometimes be unrealistic.
When $b$ is unknown, it may be reasonable to assume use of a sample-based estimate for sample allocation purposes.
Practitioners can assess sensitivity of allocations to different assumed levels of $b$, following \S\ref{subsec:ch2:simulation_misspecified_b}; if the allocations are fairly invariant to misspecified $b$, as occurred in several simulated populations, then assuming an estimated $b$ should work well.
In contrast, if allocations are sensitive to misspecified $b$, then performance of our proposed allocation methods will be unclear.
Therefore, another area of extension entails allocating sample when $b$ is unknown or is estimated using a small sample.

Another area of possible extension could entail consideration of alternative loss functions to accommodate other survey goals or even to compromise across multiple objectives.
Although we focused on design for a single objective, our methods could straightforwardly be adapted to multi-purpose surveys by constructing a multi-objective loss function as a importance-weighted combination of various single-objective loss functions, in a manner analogous to \citet{Valliant1997}; an alternative approach would be to use a fuzzy programming technique to compromise between different objectives, following \citet{Swain2016}.
Although we assumed squared error loss, \citet{Chaloner1995} and \citet{Ryan2016} provide examples of other objective functions that have been used in Bayesian experimental design.

An additional opportunity for extension is in identifying other ways to express prior knowledge, particularly when a pilot is unavailable.
Other approaches to obtaining a prior, which have been employed in the context of responsive survey design for household surveys, entail soliciting and pooling expert opinion \citep{Coffey2020} and using an intensive literature review \citep{West2021}.
Substituting one of these methods in place of pilot data would presumably entail redefining $\underset{\alpha_h,v_h|D1}{\mathrm{E}}(v_h)$ (as pertains to Equation [\ref{eqn:EV_fixed_b_final}]), which is the mechanism by which the pilot data affect the allocation.

\printbibliography

\newpage

\appendix

\counterwithin*{figure}{part}
\stepcounter{part}
\setcounter{figure}{0}
\setcounter{table}{0}

\appendixpagenumbering
\newpage \section{Application and extension of Ericson's results}

\label{app:Ericson_application_extension}

Under our problem formulation in Section 3 of the main paper, %\S\ref{sec:ch2:BayesSTSRS_analytic_results},
the model for stratum $h$ can be viewed as a special case of a regression model considered by \citet[][sec. 5.1]{Ericson1969}, insofar as we assume use of a diffuse prior and incorporate the coefficient of heteroscedasticity directly in the variance structure, whereas Ericson's model allowed for other priors and only assumed that the variance function was some known function of the auxiliary variable.
This section applies Ericson's results to our problem, and we also obtain an important distributional result that is used in our preposterior analysis.

Specific sections are as follows:
\begin{itemize}
	\item \ref{app:sec:Ericson:normalgamma} provides a definition of the normal-gamma distribution.
	\item \ref{app:EricsonPosteriorSection} applies a diffuse prior to Ericson's posterior distribution of the superpopulation parameters given the data and provides key results.
	This section primarily uses Ericson's original notation, but the results are translated back to our own notation in \ref {app:EricsonPosteriorSection:subsec2}.
	\item \ref{app:posterior_finitepopmean} provides the posterior mean and variance of $\bar{Y}|D2,e_1,b$.
	\item \ref {app:QuadFormAppendix} uses some algebraic manipulation and results relating to quadratic forms to obtain the distribution of a term that appears in the expression for $\mathrm{Var} \left(\bar{Y}_h | D2,e_1,b\right)$.
\end{itemize}
%Conduct preposterior analysis}
%subsec:Known_b_preposterior}

\label{app:sec:Ericson}
\subsection{Normal-gamma definition}\label{app:NormalGammaDef}
\label{app:sec:Ericson:normalgamma}
\begin{itemize}
	\item Suppose $X|T \sim N\left(\mu,\frac{1}{\lambda T}\right)$, where the parameters are mean and variance.
	We have a conditional pdf of $f(x|\tau)\propto \sqrt{\lambda \tau } exp\left(\frac{-(x-\mu)^2}{2} \lambda \tau\right)$.
	\item Suppose further that $T|\alpha,\beta \sim \Gamma(\alpha,\beta)$, where the parameters are shape and rate; that is, our pdf is $f(\tau)=\frac{\beta^\alpha \tau^{\alpha-1}e^{-\beta \tau}}{\Gamma(\alpha)}$
	\item Then $(X,T)\sim NormalGamma(\mu,\lambda,\alpha,\beta)$, with pdf $f(x,\tau)=\frac{\beta^\alpha\sqrt{\lambda}}{\Gamma(\alpha)\sqrt{2\pi}} \tau^{\alpha-\frac{1}{2}}e^{-\beta \tau} e^{\frac{-\tau \lambda(x-\mu)^2}{2}}$
\end{itemize}

\noindent Note that this means that $\frac{1}{T}|(\alpha,\beta)\sim IG(\alpha,\beta)$.
Via properties of the IG distribution, we have that $\mathrm{E}\left(\frac{1}{T}\right)=\frac{\beta}{\alpha-1}$ (assuming $\alpha>1$) and $\mathrm{Var}\left(\frac{1}{T}\right)=\frac{\beta^2}{(\alpha-1)^2(\alpha-2)}$ (assuming $\alpha>2$).
We also have that $\mathrm{E}(X)=\mu$ and $\mathrm{Var}(X)=\frac{\beta}{\lambda(\alpha-1)}$.

\subsection{Ericson's posterior distribution under a diffuse prior}
\label{app:EricsonPosteriorSection}

\citet{Ericson1969} provides relevant results for a univariate regression model, which we will summarize using his notation, with subsequent application to our problem.
To assist the reader, a key for selected notation is provided below:
\begin{table}[ht]
	\centering
	\caption{Notational key}
	\label{my-label}
	\begin{tabular}{lcc}
		{\ul\textbf{Description}}	&	{\ul \textbf{Ericson's Notation}} & {\ul \textbf{Our Notation}} \\
		Model	& $X_i | (y_i, \alpha, h, z_i) \overset{ind}{\sim} N \left(\alpha y_i, \frac{z_i}{h}\right)$ & $Y_{hi}|(X_{hi},\alpha_h,v_h,b) \overset{ind}{\sim} N(\alpha_h X_{hi},v_hX_{hi}^{b})$ \\
		Outcome variable	&	$X_i$ & $Y_{hi}$ \\
		Auxiliary data	&	$y_i$ & $X_{hi}$ \\Slope	&	$\alpha$ & $\alpha_h$ \\
		Variance component 1	&	$v=\frac{1}{h}$ & $v_h$ \\
		Variance component 2 &	$z_i=g(y_i)$ & $z_{hi} = x_{hi}^b$ 
	\end{tabular}
\end{table}

Ericson considered a population of $N$ distinguishable elements, labeled by the integers from the label set $\mathcal{N}=\{1,2,...,N\}$, where $X_i$ is the unknown value of an outcome variable and $y_i$ is some known and positively-valued auxiliary variable for the $i$th member of the population, defined for $i \in \mathcal{N}$.
The sample, $(s,\mathbf{x})$, comprises some set of indices of distinct population units, $s = \{ i_1,...,i_n \} \subseteq \mathcal{N}$, in conjunction with their observed values $x_j$ of $X_j$, $j\in s$.
The $X_i$'s are viewed as exchangeable and independent random variables coming from a common distribution, conditional on the superpopulation parameters.
In \S5.1, Ericson considers an unstratified model of $X_i | (y_i, \alpha, h, z_i) \overset{ind}{\sim} N \left(\alpha y_i, \frac{z_i}{h}\right)$, wherein the second parameter in the normal distribution is variance, and where $z_i=g(y_i)$, for some pre-specified and positively valued function $g$.
Given this model, he writes the joint prior for the population vector $\mathbf{X}=(X_1,X_2,...,X_N)$ in Equation (63) as:
\begin{align}
	p(\mathbf{X}|\alpha,h) \propto \prod_{i=1}^N \left(\frac{h}{z_i}\right)^{\frac{1}{2}} \mathrm{exp}\left\{-\frac{1}{2}\frac{h}{z_i}(X_i-\alpha y_i)^2\right\}   \label{eqn:Ericson:eqn63}
\end{align}
\noindent Further, Ericson assumed a normal-gamma prior on $(\alpha,h)$, written in Equation (64) as having density
\begin{align}
	f'\left(\alpha,h\right) \propto exp\left\{-\frac{1}{2} h n'(\alpha-\bar{\alpha}')^2\right\}h^{\frac{1}{2}\delta(n')} exp\left\{-\frac{1}{2} h\nu'\upsilon'\right\} h^{\frac{1}{2}\nu'-1} \label{eqn:Ericson:eqn64}
\end{align}
\noindent where $\delta(n')=0$ if $n'=0$ and $\delta(n')=1$ otherwise.
From this pdf, we can see that Ericson has assigned the parameters a distribution of $(\alpha,h)\sim NG\left(\bar{\alpha}',n',\frac{1}{2} \left[\delta(n')+\nu'-1\right],\frac{1}{2}\nu'\upsilon'\right)$, as per the normal-gamma definition in Appendix \ref{app:sec:Ericson:normalgamma}.

This leads to a posterior distribution that is normal-gamma, which he writes in Equations (67--71) as
\begin{align}
	f\left\{\alpha,h|(s,\mathbf{x})\right\} \propto exp\left\{-\frac{1}{2} h n''(\alpha-\bar{\alpha}'')^2\right\}h^{\frac{1}{2}\delta(n'')} exp\left\{-\frac{1}{2} h\nu''\upsilon''\right\} h^{\frac{1}{2}\nu''-1},
\end{align}
where
\begin{align}
	n''&=\sum_{i\in s} y_i^2/z_i+n', \label{eqn:Ericson:eqn68}
	\\ \bar{\alpha}''  &= \frac{1}{n''}\left(\sum_{i\in s} \frac{x_i y_i}{z_i} + n'\bar{\alpha}'\right) , \label{eqn:Ericson:eqn69} 
	\\ \nu'' \upsilon'' &= \left\{ \nu'\upsilon' + \frac{n'}{n''}\sum_{i\in s}\left(\frac{x_i-\bar{\alpha}'y_i}{\sqrt{z_i}}\right)^2 
	+ \frac{1}{n''}\sum_{i\in s}\frac{y_i^2}{z_i}\sum_{i\in s}\frac{x_i^2}{z_i}-\frac{1}{n''}\left(\sum_{i\in s}\frac{x_i y_i}{z_i}\right)^2\right\}, \label{eqn:Ericson:eqn70} 
	\\ \nu '' &= n+\nu' + \delta(n') - \delta(n''), \label{eqn:Ericson:eqn71}
\end{align}
and where $\delta(n'')=0$ if $n''=0$ and $\delta(n'')=1$ otherwise.
Equivalently, we have that $\alpha,h|(s,\mathbf{x})\sim NG\left(\bar{\alpha}'',n'',\frac{1}{2} \left[\delta(n'')+\nu''-1\right],\frac{1}{2}\nu''\upsilon''\right)$.
From this, we see that $h|(s,\mathbf{x})\sim \Gamma\left(\frac{1}{2} \left[\delta(n'')+\nu''-1\right],\frac{1}{2}\nu''\upsilon''\right)$, and equivalently, $\frac{1}{h}|(s,\mathbf{x})\sim IG\left(\frac{1}{2} \left[\delta(n'')+\nu''-1\right],\frac{1}{2}\nu''\upsilon''\right)$, where $IG$ refers to the inverse-gamma distribution.

Ericson noted that a diffuse prior of $f'(\alpha,h)\propto h^{-1}$ could be considered as a special case of the prior in (\ref{eqn:Ericson:eqn64}) by putting $\nu'=n'=0$.
We do so here.
Thus, Equations (\ref{eqn:Ericson:eqn68}--\ref{eqn:Ericson:eqn71}) can be simplified as:
\begin{align}
	n'' &= \sum_{i\in s} y_i^2/z_i \label{eqn:Ericson:simplified_eqn_68}
	\\ \bar{\alpha}'' &= \frac{1}{n''}\left(\sum_{i\in s} \frac{x_i y_i}{z_i} \right)  
	\\ \nu'' \upsilon'' &= \sum_{i\in s}\frac{x_i^2}{z_i}-\frac{\left(\sum_{i\in s}\frac{x_i y_i}{z_i}\right)^2}{\sum_{i\in s}\frac{y_i^2}{z_i}}
	\\ \nu '' &=  n-1 \label{eqn:Ericson:simplified_eqn_71}
	\\ \delta(n'') &= 1 
\end{align}
\noindent where $z_i=g(y_i)$.

Summing up, the posterior distribution under the diffuse prior of $f'(\alpha,h)\propto h^{-1}$ can be written as $\alpha,h|(s,\mathbf{x})\sim NG\left(\bar{\alpha}'',n'',\frac{n-1}{2},\frac{1}{2}\nu''\upsilon''\right)$, with terms as defined in (\ref{eqn:Ericson:simplified_eqn_68}--\ref{eqn:Ericson:simplified_eqn_71}) above.

\subsubsection{Posterior distribution of 1/h}
\label{app:EricsonPosteriorSection:subsec1}
From the above, and continuing to use Ericson's notation, we have that $\frac{1}{h}\sim IG\left(\frac{n-1}{2},\frac{1}{2}\nu'' \upsilon'' \right)$.
If we define $q = \frac{1}{n-3} \left(\sum_{i\in s}\frac{x_i^2}{z_i}-\frac{\left(\sum_{i\in s}\frac{x_i y_i}{z_i}\right)^2}{\sum_{i\in s}\frac{y_i^2}{z_i}}\right)$, then we have that $\frac{1}{h}\sim IG\left(\frac{n-1}{2},\frac{n-3}{2} q \right)$.

Via properties of the IG distribution, we have that if $X\sim IG(\alpha,\beta)$, then $\mathrm{E}(X)=\frac{\beta}{\alpha-1}$ (assuming $\alpha>1$) and $\mathrm{Var}(X)=\frac{\beta^2}{(\alpha-1)^2(\alpha-2)}$ (assuming $\alpha>2$).

Hence, assuming that $n>3$, we have:
\begin{align}
	\mathrm{E}\left(\frac{1}{h}\right) = q \label{eqn:Ericson_Ep_v}
\end{align}

Assuming that $n>5$, we have:
\begin{align}
	\mathrm{Var}\left(\frac{1}{h}\right) &= \frac{2q^2}{n-5}
\end{align}

\subsubsection{Ericson's posterior under our notation}\label{app:EricsonPosterior_my_notation}
\label{app:EricsonPosteriorSection:subsec2}
%\iffalse \tinytodo{Notation of this subsection is not necessarily consistent with other sections, for instance, with respect to the four NG params.} \fi

Ericson's posterior distribution for the superpopulation parameters, which were provided earlier using Ericson's notation, can be summarized in our notation as follows.
Letting $g_h=\frac{1}{v_h}$, assume a prior on $(\alpha_h, g_h)$ with density $\pi(\alpha_h, g_h)\propto \frac{1}{g_h}$; write this as $\pi(\alpha_h,\frac{1}{v_h}) \propto v_h$.
This prior, in conjunction with our model and associated assumptions, leads to a posterior distribution of $(\alpha_h,\frac{1}{v_h})|b,D2 \sim NG(\mu_h,c_h,\gamma_h,\delta_h)$ where $z_{hi}=x_{hi}^b$, $a_h=\sum_{i=1}^{n_h} \frac{y_{hi}^2}{z_{hi}}$, $d_h=\sum_{i=1}^{n_h}\frac{x_{hi}y_{hi}}{z_{hi}}$, $c_h=\sum_{i=1}^{n_h} \frac{x_{hi}^2}{z_{hi}}$, $\mu_h = \frac{d_h}{c_h}=\frac{\sum_{i=1}^{n_h}x_{hi}y_{hi}/z_{hi}}{\sum_{i=1}^{n_h} x_{hi}^2/ z_{hi}}$, $\gamma_h = \frac{n_h-1}{2}$, and $\delta_h = \frac{1}{2}\left(a_h - \frac{d_h^2}{c_h}\right)=\frac{1}{2}\left\{\sum_{i=1}^{n_h} y_{hi}^2/z_{hi} - \frac{\left[\sum_{i=1}^{n_h}x_{hi}y_{hi}/z_{hi}\right]^2}{\sum_{i=1}^{n_h}x_{hi}^2/z_{hi}}\right\}$.
Hence, we have that $v_h|b,D2 \sim IG \left(\gamma_h,\delta_h\right)$ and \\$\alpha_h|v_h,b,D2 \sim N\left(\mu_h,\frac{v_h}{c_h}\right)$.

\subsection{Posterior of finite population mean under diffuse prior}
\label{app:posterior_finitepopmean}
For the finite population mean $\mu$, then Ericson's Equations (73) and (74) with his notation are:
\begin{align}
	\mathrm{E}\left\{\mu|(s,\mathbf{x})\right\} &= \frac{n}{N}\bar{x}_s + \frac{N-n}{N}\bar{\alpha}''\bar{y}_{\bar{s}}
	\\ \mathrm{Var}\left\{\mu|(s,\mathbf{x})\right\} &= \frac{\nu''\upsilon''}{\nu''-2}\frac{\left(n''z_{\bar{s}}+y^2_{\bar{s}}\right)}{n'' N^2}
\end{align}
\noindent where the subscript $s$ denotes a sample quantity, subscript $\bar{s}$ denotes a quantity for elements not sampled, and his model is $X_i|(y_i,\alpha,h)\overset{ind}\sim N\left(\alpha y_i, \frac{z_i}{h}\right)$, and other terms are as described in Appendix \ref{app:EricsonPosteriorSection}.

Under a diffuse prior of $\pi(\alpha,h) \propto h^{-1}$, via Ericson's notation, we have:
\begin{align}
	\mathrm{E}\left\{\mu|(s,\mathbf{x})\right\} &= \frac{n}{N}\bar{x}_s + \frac{N-n}{N}\bar{\alpha}''\bar{y}_{\bar{s}}
	\nonumber \\ &= \frac{n}{N}\bar{x}_s + \frac{N-n}{N} \bar{y}_{\bar{s}}  \frac{\left(\sum_{i\in s} \frac{x_i y_i}{z_i} \right)}{\sum_{i\in s} y_i^2/z_i}
\end{align}

Translating this back to our own notation yields:
\begin{align}
	\mathrm{E}\left(\bar{Y}_h | D2,e_1,b\right) = \frac{n_h}{N_h}\bar{y}_h + \left(\frac{N_h\bar{X}_h-n_h \bar{x}_h}{N_h}\right)\left(\frac{\sum_{i\in s_{2h}} \frac{x_{hi}y_{hi}}{z_{hi}}}{\sum_{i\in s_{2h}}\frac{x_{hi}^2}{z_{hi}}}\right)
\end{align}
We can then combine results across strata to get:
\begin{align}
	\mathrm{E}\left(\bar{Y} | D2,e_1,b\right) &= \sum_{h=1}^H \frac{N_h}{N} \left\{\frac{n_h}{N_h}\bar{y}_h + \left(\frac{N_h\bar{X}_h-n_h \bar{x}_h}{N_h}\right)\left(\frac{\sum_{i\in s_{2h}} \frac{x_{hi}y_{hi}}{z_{hi}}}{\sum_{i\in s_{2h}}\frac{x_{hi}^2}{z_{hi}}}\right)\right\}
	\nonumber \\ &= \sum_{h=1}^H \frac{1}{N} \left\{n_h \bar{y}_h + \left(N_h\bar{X}_h-n_h \bar{x}_h\right)\left(\frac{\sum_{i\in s_{2h}} \frac{x_{hi}y_{hi}}{z_{hi}}}{\sum_{i\in s_{2h}}\frac{x_{hi}^2}{z_{hi}}}\right)\right\} \label{eqn:E_finitepopmean}
\end{align}

For the variance, under a diffuse prior, and via Ericson's notation, we have
\begin{align}
	\mathrm{Var}\left\{\mu|(s,\mathbf{x})\right\} &= \frac{\nu''\upsilon''}{\nu''-2}\frac{\left(n''z_{\bar{s}}+y^2_{\bar{s}}\right)}{n'' N^2}
	\nonumber \\ &= (n-3)^{-1}\left[\sum_{i\in s}\frac{x_i^2}{z_i}-\frac{\left(\sum_{i\in s}\frac{x_i y_i}{z_i}\right)^2}{\sum_{i\in s}\frac{y_i^2}{z_i}}\right]\frac{1}{N^2}
	\left[z_{\bar{s}}+\frac{y^2_{\bar{s}}}{n''}\right]
	\nonumber \\ &= (n-3)^{-1}\left[\sum_{i\in s}\frac{x_i^2}{z_i}-\frac{\left(\sum_{i\in s}\frac{x_i y_i}{z_i}\right)^2}{\sum_{i\in s}\frac{y_i^2}{z_i}}\right]\frac{1}{N^2}
	\left[z_{\bar{s}}+\frac{y^2_{\bar{s}}}{\sum_{i\in s}y_i^2/z_i}\right],
\end{align}
\noindent assuming that $n >3$.
Translating this back to our own notation yields
\begin{footnotesize}
	\begin{align}
		\mathrm{Var} \left(\bar{Y}_h | D2,e_1,b\right) = (n_h-3)^{-1} \left[\sum_{i\in s_{2h}}\frac{y_{hi}^2}{z_{hi}} - \frac{\left(\sum_{i\in s_{2h}}\frac{y_{hi}x_{hi}}{z_{hi}}\right)^2}{\sum_{i\in s_{2h}}\frac{x_{hi}^{2}}{z_{hi}}} \right]\frac{1}{N_h^2} \left[\sum_{i \in (U_h \cap s_{2h}^C)} z_{hi} + \frac{\left(N_h\bar{X}_h-n_h\bar{x}_h\right)^2}{\sum_{i\in s_{2h}} \frac{x_{hi}^{2}}{z_{hi}}}\right],
	\end{align}
\end{footnotesize}
\noindent assuming that $n_h > 3$ for all $h$, and where $s_{2h}^C$ refers to the complement of $s_{2h}$ and $U_h$ refers to the population within stratum $h$.
We then combine results across strata to get:
\begin{footnotesize}
	\begin{align}
		\mathrm{Var}\left(\bar{Y} | D2,e_1,b\right) &= \sum_{h=1}^H \frac{1}{N^2}\left\{\frac{1}{n_h-3} \left[\sum_{i\in s_{2h}}\frac{y_{hi}^2}{z_{hi}} - \frac{\left(\sum_{i\in s_{2h}}\frac{y_{hi}x_{hi}}{z_{hi}}\right)^2}{\sum_{i\in s_{2h}}\frac{x_{hi}^2}{z_{hi}}} \right] \left[\sum_{i \in (U_h \cap s_{2h}^C)} z_{hi} + \frac{\left(N_h\bar{X}_h-n_h\bar{x}_h\right)^2}{\sum_{i\in s_{2h}} \frac{x_{hi}^2}{z_{hi}}}\right]\right\} \label{eqn:Var_finitepopmean}
	\end{align}
\end{footnotesize}

\subsection{Quadratic form distribution}
\label{app:QuadFormAppendix}
This subsection uses results relating to quadratic forms to obtain the distribution of \\$q_{bh} = \left[\sum_{i\in s_{2h}}\frac{y_{hi}^2}{z_{hi}} - \frac{\left(\sum_{i\in s_{2h}}\frac{y_{hi}x_{hi}}{z_{hi}}\right)^2}{\sum_{i\in s_{2h}}\frac{x_{hi}^2}{z_{hi}}} \right]$, which is a quadratic form in $y$.
Specifically, we will show that $v_h^{-1} q_{bh} | \left(v_h, \alpha_h, b, s_{2h} \right) \sim \chi^2_{n_h-1}$.
As per previous notation and set-up, $\left(v_h, \alpha_h, b \right)$ are model parameters, and $s_{2h}$ is the set of sample indicators in stratum $h$.

To simplify notation, temporarily focus on a single stratum, ignore the stratum subscripts, and write the set of sample indicators as $s$.
For the remainder of this subsection, assume that the model parameters, $\left(b,v, \alpha \right)$, are known.
Also suppose that $s$ is fixed; given that the population auxiliary data, $\left\{X_{i}:i=1,...,N\right\}$ are known, it follows that the sample auxiliary data, $\left\{x_i:i=1,...,n\right\}$, are known.
We have that $y_i\sim N(\alpha x_i,v z_i)$, where $z_i = x_i^b$.
Let $a_i = \frac{y_i}{\sqrt{z_i}}$ and $d_i=\frac{x_i}{\sqrt{z_i}}$.
Also define the vectors $a = (a_1,...,a_n)^t$ and $d=(d_1,...,d_n)^t$.
Note that $a$ is random with respect to the model, whereas $d$ is fixed.
We have:
\begin{align}
	\sum_i y_{i}^2/z_{i} - \frac{\left(\sum_{i}y_{i}x_{i}/z_i \right)^2}{\sum_{i}x_{i}^{2}/z_i} &= \sum_i a_i^2 - \frac{\left(\sum_i a_i d_i\right)^2}{\sum_k d_k^2}
	\nonumber \\ &= \sum_i a_i^2 - \frac{\sum_i a_i^2 d_i^2 + \sum_{i\ne j} a_i d_i a_j d_j }{\sum_k d_k^2}
	\nonumber \\ &= \sum_i a_i^2 \left(1 - \frac{d_i^2}{\sum_k d_k^2}\right) + \sum_{i=1}^{n}\sum_{j=1,j\ne i}^n a_i a_j \left(\frac{-d_i d_j}{\sum_k d_k^2}\right)
	\nonumber \\ &= a^t D a
\end{align}

\noindent where $D$ is an $n \times n$ matrix, in which cell $D_{ij}$ in row $i$ and column $j$ is defined as:
\begin{align}
	D_{ii}&=1-\frac{d_i^2}{\sum_k d_k^2}, \tab i=1,...,n
	\\D_{ij} &= \frac{-d_i d_j}{\sum_k d^2_k}, \tab i=1,...,n;j=1,...,n;i\ne j
\end{align}

Let $D'=D^2$.
(Hence, we have $d_{rc}'=\sum_{i=1}^n d_{ri}d_{ic}$.) Then examining the diagonals, we have:
\begin{align}
	D_{ii}' &= D_{ii}^2 +  \sum_{j=1,j\ne i}^n D_{ij} D_{ji}
	\nonumber \\&= \left(1-\frac{d_i^2}{\sum_k d_k^2}\right)^2 + \frac{d_i^2}{\left(\sum_k d_k^2\right)^2} \sum_{j=1,j\ne i}^n d_j^2
	\nonumber \\&= \left(1-\frac{d_i^2}{\sum_k d_k^2}\right)^2 + \frac{d_i^2}{\left(\sum_k d_k^2\right)^2} \left(\left[\sum_{j=1}^n d_j^2\right] - d_i^2\right)
	\nonumber \\&= \left(1-\frac{d_i^2}{\sum_k d_k^2}\right)^2 + \frac{d_i^2}{\sum_k d_k^2}\left(1 - \frac{d_i^2}{\sum_k d_k^2}\right) 
	\nonumber \\&= 1-\frac{d_i^2}{\sum_k d_k^2} = D_{ii}
\end{align}

For cell $ij$ where $i\ne j$, we have:
\begin{align}
	D_{ij}' &= \sum_{k=1}^n D_{ik} D_{kj}
	\nonumber \\ &= \frac{1}{\left(\sum_k d_k^2\right)^2 }\sum_{k=1}^n d_k^2 d_i d_j 
	\nonumber \\ &= \frac{d_i d_j}{\sum_k d_k^2} = D_{ij}
\end{align}

Next, note that a general result regarding quadratic forms is that if $Y_1,...,Y_n$ are independent normal random variables where $Y_i$ has mean $\mu_i$ and known variance $\sigma^2$, and matrix $A$ is symmetric and idempotent and has dimension $n \times n$, then random variable $\frac{Y^t A Y}{\sigma^2}$ is distributed as non-central chi-square with $tr(A)$ degrees of freedom and non-centrality parameter of $\lambda=\frac{\mu^t A \mu}{\sigma^2}$, where $\mu = (\mu_1,...,\mu_n)^t$.

In our case, we have that $a_i \sim N\left(\frac{\alpha x_i}{\sqrt{z_i}} ,v \right)$, for $i=1,...,n$, with $a_1,...,a_n$ independent.
Likewise, we have that $\frac{a_i}{\alpha}\sim N(d_i, v \alpha^{-2})$, for $i=1,...,n$.
Given that $D$ is symmetric and idempotent with rank $n-1$, it follows that $\frac{\left(\frac{a}{\alpha}\right)^t D \left(\frac{a}{\alpha}\right) }{v \alpha^{-2}}=\frac{a^t D a}{v}\sim \chi^2$ with $df=n-1$ and non-centrality parameter of $\lambda = \frac{\left(\mathrm{E}\left[\frac{a}{\alpha}\right]\right)^t D \left(\mathrm{E}\left[\frac{a}{\alpha}\right]\right) }{v \alpha^{-2}}= \frac{d^t D d}{v \alpha^{-2}}$.

Note that matrix $D$ can be rewritten as $D=I - \frac{d d^t}{d^t d}$ where $d=(d_1,...,d_n)^t$ and $I$ is an $n \times n$ identity matrix.
Hence, we have that $d^t D d = d^t \left(I - \frac{d d^t }{d^t d}\right) d = d^t I d - \frac{d^t d d^t d}{d^t d} = 0$, and likewise, $\lambda=0$.
Thus, we have that $\frac{a^t D a}{v}\sim \chi^2_{n-1}$.

From here, we see that $\mathrm{E}(a^t D a) = v \mathrm{E}(\chi^2_{n-1}) = v(n-1)$.

Note that the above used the abbreviated notation, the parameters were treated as fixed, and the sample indicators were treated as fixed.
Using the more detailed notation, we have shown that $v_h^{-1} q_{bh} | \left(v_h, \alpha_h, b, s_{2h} \right) \sim \chi^2_{n_h-1}$ and that $\underset{D2|\alpha_h,v_h,b,s_{2h}}{\mathrm{E}}\left(q_{bh}\right)=v_h (n_h-1)$.

\appendixpagenumbering
\newpage
\section{Analysis for fixed and known b}
\label{app:fixed_known_b_analysis}

Recall from Equation %(\ref{eqn:Var_mainstudy}) 
(12) of the main paper that:
\begin{footnotesize}
	\begin{align}
		\mathrm{Var}\left(\bar{Y} | D2,e_1,b\right) &= \sum_{h=1}^H \frac{1}{N^2}\left\{\frac{1}{n_h-3} \left[\sum_{i\in s_{2h}}\frac{y_{hi}^2}{z_{hi}} - \frac{\left(\sum_{i\in s_{2h}}\frac{y_{hi}x_{hi}}{z_{hi}}\right)^2}{\sum_{i\in s_{2h}}\frac{x_{hi}^2}{z_{hi}}} \right] \left[\sum_{i \in (U_h \cap s_{2h}^C)} z_{hi} + \frac{\left(N_h\bar{X}_h-n_h\bar{x}_h\right)^2}{\sum_{i\in s_{2h}} \frac{x_{hi}^2}{z_{hi}}}\right]\right\}
	\end{align}
\end{footnotesize}

Let $q_{bh}=\left[\sum_{i\in s_{2h}}\frac{y_{hi}^2}{z_{hi}} - \frac{\left(\sum_{i\in s_{2h}}\frac{y_{hi}x_{hi}}{z_{hi}}\right)^2}{\sum_{i\in s_{2h}}\frac{x_{hi}^2}{z_{hi}}} \right]$ and $k(s_{2h})=\left[\sum_{i \in (U_h \cap s_{2h}^C)} z_{hi} + \frac{\left(N_h\bar{X}_h-n_h\bar{x}_h\right)^2}{\sum_{i\in s_{2h}} \frac{x_{hi}^2}{z_{hi}}}\right]$.
\\(This notation is used given that the first quantity is a quadratic form and the second quantity is fixed given the sample indicators, as the population-level auxiliary data are known.) Then we have:
\begin{align}
	\mathrm{Var}\left(\bar{Y} | D2,e_1,b\right) &= \sum_{h=1}^H \frac{1}{N^2(n_h-3)} q_{bh} k(s_{2h}) \label{eqn:v_ybar_given_d2}
\end{align}

\subsection{Analysis relating to \texorpdfstring{$q_{bh}$}{q[bh]}}
Conditional on the pilot data (D1) and the main study sample allocation ($e_1$), Equation (\ref{eqn:v_ybar_given_d2}) has three sources of uncertainty, in relation to: (1) the posterior distribution for the superpopulation parameters conditional on the pilot data; (2) the outcome variable, conditional on the superpopulation parameters and sample indicators; and (3) the set of sample indicators (which we refer to as the `design-based sample').
We will analyze $q_{bh}$ and $k(s_{2h})$ separately since it turns out that these terms are independent, which will result in $\mathrm{E}(q_{bh}k(s_{2h})) = \mathrm{E}(q_{bh})\mathrm{E}(k(s_{2h}))$.
These terms are independent since it turns out that $q_{bh}$ does not depend on the units selected in the design-based sample, $s_{2h}$, whereas $k(s_{2h})$ only depends on the design-based sample.

Thus, as a first step, we will evaluate $\underset{D2|D1}{\mathrm{E}} (q_{bh})$ through a series of conditional expectations.
For now, treating the design-based sample as fixed, we will use subscripts of $M$ and $P$, respectively, to distinguish whether the expectation is respect to the model or with respect to the posterior distribution of the parameters given the pilot data.
%(we will consider variability with respect to the design-based sample in \ref{app:FirstOrderTS_for_k}).
Let $\mathrm{E}_M(Z)=\underset{D2|\alpha_h,v_h,D1,s_{2h}}{\mathrm{E}}(Z)$ and $\mathrm{E}_P(Z)=\underset{\alpha_h,v_h|D1}{\mathrm{E}}(Z)$ for any $Z$.
Then assuming that the strata sample sizes, $\{n_h\}$, are fixed, we have:
\begin{align}
	\underset{D2|D1}{\mathrm{E}}(q_{bh}|s_{2h})=\mathrm{E}_P(\mathrm{E}_M(q_{bh}|s_{2h})) \label{eqn:EPEM_s2bh}
\end{align}

Appendix \ref{app:QuadFormAppendix} shows that $v_h^{-1}q_{bh}|(v_h,\alpha_h,b,s_{2h})\sim \chi^2_{n_h-1}$.
Note that the resulting distribution is free of $s_{2h}$, resulting in $v_h^{-1}q_{bh}|(v_h,\alpha_h,b)\sim \chi^2_{n_h-1}$.
Since the distribution $(\alpha_h,v_h)|D1$ also does not depend on $s_{2h}$, it follows that the distribution of $q_{bh}|D1$ does not depend on the main sample, and is independent from $k(s_{2h})$.
As a result of this independence, we have that
\begin{align}
	\underset{D2|D1}{\mathrm{E}} \mathrm{Var}\left(\bar{Y} | D2,e_1,b\right) &= \sum_{h=1}^H \frac{1}{N^2(n_h-3)} \left[\underset{D2|D1}{\mathrm{E}}(q_{bh}) \right] \mathrm{E}\left(k(s_{2h})\right) \label{eqn:ED3givenD2_YbargivenB}
\end{align}

In addition, from the properties of the $\chi^2$ distribution, we have that $\mathrm{E}_M(q_{bh}|s_{2h})=v_h(n_h-1)$.
Substituting this into (\ref{eqn:EPEM_s2bh}), and then using the fact that $q_{bh}|D1$ does not depend on $s_{2h}$, yields:
\begin{align}
	\underset{D2|D1}{\mathrm{E}}(q_{bh})&=(n_h-1) \mathrm{E}_P(v_h)
\end{align}

Substituting this into (\ref{eqn:ED3givenD2_YbargivenB}) gives:
\begin{align}
	\underset{D2|D1}{\mathrm{E}} \mathrm{Var}\left(\bar{Y} | D2,e_1,b\right) &= \sum_{h=1}^H \frac{1}{N^2}\frac{n_h-1}{n_h-3} \left[\mathrm{E}_P (v_h)\right] \mathrm{E}\left(k(s_{2h})\right)\label{eqn:EV_result_chisq}
\end{align}

\subsection{First-order Taylor series approximation in relation to \texorpdfstring{$k(s_{2h})$}{k(s2h)}}
\label{app:FirstOrderTS_for_k}

In the above equation, $\mathrm{E}_P(v_{h})=\underset{\alpha_h,v_h|D1}{\mathrm{E}}(v_{h})$ can be computed at the time of sampling for the main study based on the posterior distribution for the parameters given the pilot data.
The term $k(s_{2h})$ is a nonlinear function of the main study sample, which has not yet been collected.
Thus, we obtain a first-order Taylor series approximation of $\mathrm{E}(k(s_{2h}))$ so that our expected posterior loss is in terms of known population characteristics.

For most of this section, we will temporarily omit the stratum and sampling phase subscripts, in order to simplify notation.
Recall that a simple random sample (within stratum) is assumed.
Recall also that $k(s)=\sum_{i \in ns} z_{i} + \frac{(N\bar{X}-n\bar{x})^2}{\sum_{i\in s} x_{i}^2 / z_{i}}$, where $z_i=x_i^b$ and where $ns$ refers to the nonsampled units.
For the lefthand term, we have that $\mathrm{E}\left(\sum_{i\in ns}z_i\right) = (N-n) \bar{Z}$ where $Z_i=X_i^b$ and $\bar{Z}=\frac{1}{N}\sum_{i=1}^{N}Z_i$.
For the righthand term, assuming that $n \to \infty$ and $N \to \infty$, we can reasonably use a first-order Taylor series approximation, wherein
\begin{align}
	\mathrm{E}\left[\frac{\left(N \bar{X} - n \bar{x} \right)^2}{\sum_{i\in s}x^2_{i}/z_{i}} \right] 
	&\approx \frac{\mathrm{E}\left[\left(N\bar{X}-n\bar{x}\right)^2\right]}{\mathrm{E}\left(\sum_{i \in s} x^2_{i}/z_{i}\right)},
\end{align}
and where the the only source of uncertainty is with respect to the design (noting our previous assumptions that the $\{X_i\}$ and $b$ are known and are treated as fixed).

For the numerator, we have
\begin{align}
	\mathrm{E}\left[\left(N\bar{X}-n\bar{x}\right)^2\right] &= \mathrm{Var}\left[N\bar{X}-n\bar{x}\right]  + \left(\mathrm{E}\left[N\bar{X}-n\bar{x}\right]\right) ^2
	\nonumber \\ &= \mathrm{Var}(n \bar{x}) + (N-n)^2 \bar{X}^2
	\nonumber \\ &= n^2 \left(1-\frac{n}{N}\right) \frac{S^2_{x}}{n} + (N-n)^2 \bar{X}^2
	\nonumber \\ &= (N-n)\left[\frac{n}{N} S^2_{x} + (N-n) \bar{X}^2\right],
\end{align}
\noindent where $S^2_x = \frac{1}{N-1} \sum_{i=1}^N (X_i-\bar{X})^2$.
For the denominator, if we let $Q_{i}=\frac{X_{i}}{\sqrt{Z_{i}}}$, $\bar{Q}=\frac{1}{N}\sum_{i=1}^{N} Q_{i}$, $S^2_{q}=\frac{1}{N-1}\sum_{i=1}^{N}\left(Q_{i}-\bar{Q}\right)^2$, and $CV_{q} = \frac{S_{q}}{\bar{Q}}$, then we find
\begin{align}
	\mathrm{E}\left(\sum_{i \in s} x^2_{i}/z_{i}\right) &= \frac{n}{N}\sum_{i\in U} Q_{i}^2
	\nonumber \\ &= \frac{n}{N} \left((N-1)S_{q}^2 + N \bar{Q}^2 \right)
	\nonumber \\ &\approx n (S_{q}^2 + \bar{Q}^2)
	\\ &= n \bar{Q}^2 (1 + CV_{q}^2)
\end{align}

Hence, we have a first-order Taylor series approximation of
\begin{align}
	\mathrm{E}\left(k(s)\right) &\approx (N-n)\bar{Z} + \frac{(N-n)\left[\frac{n}{N} S^2_{x} + (N-n) \bar{X}^2\right]}{n \bar{Q}^2 (1 + CV_{q}^2)}
	\\ &= \left(N-n\right) \left[\bar{Z}+ \frac{\frac{n}{N} S^2_{x} + (N-n) \bar{X}^2}{n \bar{Q}^2 (1 + CV_{q}^2)}\right].
\end{align}

Reintroducing strata subscripts, we substitute this quantity into Equation (\ref{eqn:EV_result_chisq}) to yield
\begin{align}
	\underset{D2|D1}{\mathrm{E}} \mathrm{Var}\left(\bar{Y} | D2,e_1,b\right) &\approx \sum_{h=1}^H \frac{\mathrm{E}_P (v_h)}{N^2} \left(\frac{n_h-1}{n_h-3}\right) \left(N_h-n_h\right) \left[\bar{Z}_h+ \frac{\frac{n_h}{N_h} S^2_{xh} + (N_h-n_h) \bar{X}_h^2}{n_h \bar{Q}_h^2 (1 + CV_{qh}^2)}\right] \label{eqn:EV_Ybar_knownB_final},
\end{align}
which assumes large $\{n_h\}$.

\subsection{Expected value of \texorpdfstring{$v_h$}{v\_h} with respect to posterior given pilot data}
\label{app:subsec:Ep_vh}
In the above equation, $\mathrm{E}_P(v_h)=\underset{\alpha_h,v_h|D1}{\mathrm{E}}(v_h)$ can be obtained using the posterior distribution from the pilot data.
This could be easily estimated via standard Bayesian computation.
Alternatively, if we use the pilot data in conjunction with a diffuse prior on the model parameters then we can draw upon results from Appendix \ref{app:EricsonPosteriorSection}, which provides basic results of Ericson's posterior distribution under the use of a diffuse prior.
Under such a diffuse prior of $\pi\left(\alpha_h,\frac{1}{v_h}\right) \propto v_h$, then translating the notation of Equation (\ref{eqn:Ericson_Ep_v}) to the current notation yields:
\begin{align}
	\mathrm{E}_P(v_h) &= \frac{1}{m_h-3} \left(\sum_{i\in s{1h}}\frac{y_{hi}^2}{z_{hi}}-\frac{\left(\sum_{i\in s_{1h} }\frac{y_{hi} x_{hi}}{z_{hi}}\right)^2}{\sum_{i\in s_{1h}}\frac{x_{hi}^2}{z_{hi}}}\right) \label{eqn:ep_vh}
\end{align}
\noindent where, as before, $m_h$ is the pilot sample size in stratum $h$, $s_{1h}$ is the set of units in the pilot sample in stratum $h$, and $z_{hi}=x_{hi}^b$.

\subsection{Sampling-based approximation for \texorpdfstring{$k(s_{2h})$}{k(s2h)}}
\label{app:MC_approx_for_ks2h}

This section outlines an efficient Monte Carlo (MC) sampling-based method to approximating $\mathrm{E}(k(s_{2h}))$, as an alternative to the TS approximation of Appendix \ref{app:FirstOrderTS_for_k}.
This MC approach can be incorporated directly into the optimization process, while avoiding a potential computational bottleneck that could arise from a naive implementation.

Recall that $k(s_{2h})=\left[\sum_{i \in (U_h \cap s_{2h}^C)} z_{hi} + \frac{\left(N_h\bar{X}_h-n_h\bar{x}_h\right)^2}{\sum_{i\in s_{2h}} \frac{x_{hi}^2}{z_{hi}}}\right]$, where $Z_{hi}=X_{hi}^b$ and $b$ is known.
For purposes of this section, define $j_h = \frac{\left(N_h\bar{X}_h-n_h\bar{x}_h\right)^2}{\sum_{i\in s_{2h}}\frac{x^2_{hi}}{z_{hi}}}$.
We have that
\begin{align}
	\mathrm{E}\left(k(s_{2h}|n_h)\right) = (N_h-n_h)\bar{Z}_h + \mathrm{E}(j_h|n_h). \label{eqn:paper1_appendix_jrn}
\end{align}
For optimization purposes, we will need to approximately evaluate $\mathrm{E}(j_h|n_h)$ at numerous possible allocations.
For a given stratum, consider $R$ random permutations of elements for large $R$ (e.g., 10,000).
For a given permutation, treat the first $n_h$ units of stratum $h$ as an SRS of size $n_h$, and compute a sample estimate for every possible value of $n_h$, using cumulative sums.
This will yield the vector $\left\{\hat{\mathrm{E}}(j_h|n_h):n_h=1,2,...,N_h\right\}$, in advance of optimization.

More specifically, consider a single stratum, and ignore strata subscripts to simplify notation.
Let $(i_{r1}, i_{r2},...,i_{rN})$ refer to the $r$th random permutation of the population indices, for $r=1,2,...,R$.
For the $r$th permutation, compute $\mathbf{j}_r = (j_{r1},j_{r2},...,j_{rN})$, where we define
\begin{align}
	j_{rn}=\frac{(N\bar{X}-\sum_{j=1}^n X_{i_{rj}})^2}{\sum_{j=1}^n\frac{X^2_{i_{rj}}}{Z_{i_{rj}}}},
\end{align}
and where the summations of $\{X_i\}$ and $\left\{\frac{X_i^2}{Z_i}\right\}$ are computed at all levels via cumulative sums, as to share computation across different sample sizes.
Our MC estimator is thus
\begin{align}
	\hat{\mathrm{E}}(k(s_2)|n)=(N-n)\bar{Z}+\bar{j}_{.n}, \label{eqn:paper1_Ehat_ks2}
\end{align}
where $\bar{j}_{.n}=\frac{1}{R}\sum_{r=1}^R j_{rn}$.
The MC estimate of (\ref{eqn:paper1_Ehat_ks2}) will reflect the sample mean of $R$ i.i.d.\ random variables, $j_{h1}, ..., j_{hR}$, reflecting a common distribution, $f(j_{hi}|n_h)$; we suggest choosing $R$ adequately high so the CV of (\ref{eqn:paper1_Ehat_ks2}), which can be estimated empirically, is small enough across potential sample sizes such that choosing larger $R$ would not meaningfully alter the resulting allocations.
Repeating this process separately for each stratum and then reintroducing strata subscripts will yield $H$ vectors of Monte Carlo estimates of $\{k(s_{2h}|n_h):n_h=1,...,N_h\}$.
These vectors are precomputed in advance of optimization with large $R$, subsequently allowing for quick and accurate evaluation for any given integer sample allocation.
%Considering that $\bar{j}_{.n}$ is the sample mean of $R$ i.i.d. random variables, we have that $SD(\bar{j}_{.n}) = \frac{SD(\bar{j}_{rn})}{\sqrt{R}}$; $R$ should be chosen adequately large so that the CV of (\ref{eqn:paper1_Ehat_ks2}) is adequately small across feasible allocations.

The quantity $k(s_{2h})$ is only defined for discrete $\{n_h\}$, but many optimization methods assume that decision variables are continuous.
This disconnect can be accommodated straightforwardly through linear interpolation, for example, obtain the precomputed MC estimates for stratum $h$ sample sizes of $\lfloor n_h \rfloor$ and $\lceil n_h \rceil$, then compute a weighted average via
\begin{align}
	\hat{\mathrm{E}}(k(s_{2h})|n_h) = \left(1 - frac_h\right)\hat{\mathrm{E}}(k(s_{2h})|\lfloor n_h \rfloor) + \left(frac_h\right) \hat{\mathrm{E}}(k(s_{2h})|\lceil n_h \rceil),
\end{align}
where $frac_h = n_h - \lfloor n_h \rfloor$ denotes the fractional allocation component.
This step should enable continuous optimization algorithms to yield reasonable approximations of the optimum allocation in scenarios where the fractional component of the allocation does not have much impact (e.g., for large strata sample sizes).
If a more exact situation is needed, integer programming could be employed in the neighborhood of the continuous solution.

\appendixpagenumbering
\newpage
\section{Additional simulation detail}

\subsection{Selection of \texorpdfstring{$v_h$}{v\_h} to achieve target correlation}
\label{app:paper1_target_correlation_explanation}

Our simulation (from Sections 4 and 5 of the main paper)
%\S\ref{sec:ch2:Simulation_design} and \S\ref{sec:ch2:sim_results}) 
involved generating pseudo-populations that followed our model for a given set of size measures, $\{X_{hi}\}$, and model parameters, $\{b,\alpha_h,v_h\}$.
%\section{Simulation design} = \label{sec:ch2:Simulation_design}
%\section{Simulation results} = \label{sec:ch2:sim_results}
In doing so, after directly specifying $b$ and $\{\alpha_h\}$, we specified target correlations of $\{\rho_h\}$ and then determined $\{v_h\}$ via Equation (14) of the main paper %(\ref{eqn:vh_given_rhoh}) 
so as to approximately result in the desired correlations between $\{X_{hi},Y_{hi}\}$ for the $H$ strata.

Equation (14) of the main paper %(\ref{eqn:vh_given_rhoh}) 
was derived as follows: To simplify notation, ignore strata subscripts.
Define $\rho_{xy}=\frac{C_{xy}}{S_x S_y}$, $S_x^2 = \frac{1}{N-1}\sum_{i=1}^{N} (X_i-\bar{X})^2$, $S_y^2 = \frac{1}{N-1}\sum_{i=1}^{N} (Y_i-\bar{Y})^2$, and $C_{xy} = \frac{1}{N-1}\sum_{i=1}^{N} (X_i-\bar{X})(Y_i-\bar{Y})$, wherein $\bar{X}=N^{-1}\sum_{i=1}^N X_i$ and $\bar{Y}=N^{-1}\sum_{i=1}^N Y_i$.
We aim to specify $v$ such that $\mathrm{E}\left(\rho_{xy}\right) \approx \rho_{xy}^0$ for some specified $\rho_{xy}^0$, and where variability is considered with respect to the distribution for $\{Y_i\}|(\{X_i\}, \alpha, v, b)$.
We find that $\mathrm{E}(C_{xy}) = \alpha S_x^2$ and $\mathrm{E}(S_y^2) = \alpha^2 S_x^2 + v \bar{Z}$, where $\bar{Z}= N^{-1}\sum_{i=1}^N X_i^b$.
Assuming that $\mathrm{E}(\rho_{XY}) \approx \frac{\mathrm{E}(C_{xy})}{\mathrm{E}(S_x S_y)}$ and $\mathrm{E}(S_y) \approx \sqrt{\mathrm{E}(S^2_y)}$, we have that $\mathrm{E}(\rho_h) \approx \frac{\alpha S_x}{\sqrt{\alpha^2 S_x^2 + v\bar{Z}}}$.
Solving for $v$ and reintroducing strata subscripts leads to an approximate result analogous to Equation (14) of the main paper, %(\ref{eqn:vh_given_rhoh}),
meaning that the data-generating mechanism described earlier would approximately yield the desired correlations, on expectation, if the assumptions hold.

 \subsection{Supplementary figures and tables}

\begin{landscape}
	\begin{singlespace}
		\begin{table}

\caption{\label{tab:RelBias_by_metric}1,000*RelBias by main strategy and population-generating characteristics}
\resizebox{\linewidth}{!}{
\begin{tabular}[t]{l>{}l|rrrrr>{}r|rrrrr>{}r|rrrrr>{}r|}
\toprule
\multicolumn{1}{c}{ } & \multicolumn{1}{c}{ } & \multicolumn{6}{c}{MOS1 (most skewness)} & \multicolumn{6}{c}{MOS2 (mid skewness)} & \multicolumn{6}{c}{MOS3 (least skewness)} \\
\cmidrule(l{3pt}r{3pt}){3-8} \cmidrule(l{3pt}r{3pt}){9-14} \cmidrule(l{3pt}r{3pt}){15-20}
\rotatebox{90}{ } & \rotatebox{90}{ } & \rotatebox{90}{1. Baseline} & \rotatebox{90}{2. Lower corrs} & \rotatebox{90}{3. Higher corrs} & \rotatebox{90}{4. Inc. corrs} & \rotatebox{90}{5. Dec. slopes} & \rotatebox{90}{6. Inc. corrs, dec. slopes} & \rotatebox{90}{1. Baseline} & \rotatebox{90}{2. Lower corrs} & \rotatebox{90}{3. Higher corrs} & \rotatebox{90}{4. Inc. corrs} & \rotatebox{90}{5. Dec. slopes} & \rotatebox{90}{6. Inc. corrs, dec. slopes} & \rotatebox{90}{1. Baseline} & \rotatebox{90}{2. Lower corrs} & \rotatebox{90}{3. Higher corrs} & \rotatebox{90}{4. Inc. corrs} & \rotatebox{90}{5. Dec. slopes} & \rotatebox{90}{6. Inc. corrs, dec. slopes}\\
\midrule
 & b = 0 & -0.2 & 0.3 & -0.2 & 0.6 & 0.3 & 0.5 & 0.1 & -0.3 & -0.2 & -0.3 & -0.4 & -0.2 & 0.7 & -0.2 & -0.1 & 0.2 & 0.0 & -0.4\\

 & b = 0.5 & 0.7 & 0.5 & 0.1 & 0.6 & -0.9 & -0.0 & 0.1 & 0.1 & 0.2 & 0.1 & 0.3 & 0.4 & -0.4 & 0.2 & -0.1 & 0.3 & -0.2 & 0.3\\

 & b = 1 & 0.0 & -0.3 & 0.6 & 1.0 & -0.8 & -0.3 & -0.0 & 0.1 & -0.1 & -0.5 & 0.1 & -0.6 & -0.2 & -0.4 & -0.0 & -0.1 & -0.1 & -0.4\\

 & b = 1.5 & -0.6 & 0.4 & 0.4 & 1.1 & 0.6 & 0.7 & 0.2 & 0.1 & -0.1 & -0.3 & 0.1 & 0.1 & -0.2 & -0.0 & -0.3 & -0.3 & 0.3 & 0.1\\

\multirow{-5}{*}{\raggedright\arraybackslash N-HT} & b = 2 & 0.1 & 0.9 & 0.1 & 0.6 & -0.9 & -0.3 & 0.4 & -0.7 & -0.0 & 0.4 & 0.3 & -0.3 & 0.2 & -0.5 & 0.2 & -0.1 & -0.0 & 0.2\\
\cmidrule{1-20}
 & b = 0 & -1.2 & -0.3 & -0.0 & 0.0 & -0.6 & 0.2 & -0.5 & 0.1 & -0.0 & 0.5 & -0.1 & 0.4 & -0.2 & 0.3 & -0.1 & -0.2 & -0.5 & -0.2\\

 & b = 0.5 & -0.1 & -0.8 & -0.2 & -0.5 & -0.3 & 0.9 & 0.3 & -0.4 & 0.1 & -0.7 & -0.1 & -0.3 & 0.0 & 0.2 & 0.0 & 0.1 & 0.1 & -0.2\\

 & b = 1 & -0.3 & 0.4 & -0.2 & -0.4 & -0.2 & -0.1 & -0.1 & 0.1 & 0.2 & 0.1 & 0.2 & 0.3 & -0.1 & 0.2 & 0.1 & 0.2 & -0.1 & 0.7\\

 & b = 1.5 & 0.1 & -0.0 & 0.0 & -0.5 & -0.1 & -0.7 & 0.1 & -0.3 & 0.0 & -0.1 & -0.1 & 0.8 & -0.0 & 0.1 & -0.3 & -0.0 & 0.4 & 0.1\\

\multirow{-5}{*}{\raggedright\arraybackslash C-SR} & b = 2 & 0.1 & 0.2 & 0.2 & 0.3 & 0.2 & 0.9 & 0.0 & -1.4 & 0.1 & -0.3 & 0.2 & -0.4 & -0.3 & 0.2 & 0.1 & -0.0 & -0.2 & 0.1\\
\cmidrule{1-20}
 & b = 0 & 1.0 & 1.1 & 0.4 & -1.5 & 0.2 & 2.4 & 0.0 & 0.9 & 0.1 & -0.0 & -0.2 & 1.6 & 0.3 & -0.6 & -0.1 & 0.6 & 0.3 & 0.5\\

 & b = 0.5 & 1.1 & 0.8 & 0.0 & -0.7 & -0.3 & 1.3 & -0.0 & 0.6 & -0.2 & -0.8 & -0.4 & -0.2 & 0.4 & 0.0 & 0.1 & -0.1 & -0.3 & 0.9\\

 & b = 1 & -0.2 & -0.4 & -0.1 & 0.0 & 0.0 & 1.1 & -0.3 & 0.2 & -0.2 & -0.4 & -0.2 & 0.6 & 0.1 & 0.4 & 0.0 & -0.3 & 0.3 & -0.1\\

 & b = 1.5 & -0.4 & -0.3 & -0.1 & 1.0 & -0.3 & -0.8 & -0.3 & 0.5 & 0.1 & 0.3 & 0.1 & 0.1 & 0.2 & -0.4 & 0.1 & 0.1 & -0.3 & -0.4\\

\multirow{-5}{*}{\raggedright\arraybackslash B-P} & b = 2 & -0.4 & 2.3 & 0.6 & 2.9 & 1.6 & 2.4 & -0.4 & 1.5 & 0.2 & -1.1 & 1.2 & 0.7 & 0.5 & -0.6 & -0.0 & -1.1 & 0.7 & 0.4\\
\bottomrule
\end{tabular}}
\end{table}

		\begin{table}

\caption{\label{tab:Coverage_by_metric}CI coverage rate (\%) by main strategy and population-generating characteristics}
\resizebox{\linewidth}{!}{
\begin{tabular}[t]{l>{}l|rrrrr>{}r|rrrrr>{}r|rrrrr>{}r|}
\toprule
\multicolumn{1}{c}{ } & \multicolumn{1}{c}{ } & \multicolumn{6}{c}{MOS1 (most skewness)} & \multicolumn{6}{c}{MOS2 (mid skewness)} & \multicolumn{6}{c}{MOS3 (least skewness)} \\
\cmidrule(l{3pt}r{3pt}){3-8} \cmidrule(l{3pt}r{3pt}){9-14} \cmidrule(l{3pt}r{3pt}){15-20}
\rotatebox{90}{ } & \rotatebox{90}{ } & \rotatebox{90}{1. Baseline} & \rotatebox{90}{2. Lower corrs} & \rotatebox{90}{3. Higher corrs} & \rotatebox{90}{4. Inc. corrs} & \rotatebox{90}{5. Dec. slopes} & \rotatebox{90}{6. Inc. corrs, dec. slopes} & \rotatebox{90}{1. Baseline} & \rotatebox{90}{2. Lower corrs} & \rotatebox{90}{3. Higher corrs} & \rotatebox{90}{4. Inc. corrs} & \rotatebox{90}{5. Dec. slopes} & \rotatebox{90}{6. Inc. corrs, dec. slopes} & \rotatebox{90}{1. Baseline} & \rotatebox{90}{2. Lower corrs} & \rotatebox{90}{3. Higher corrs} & \rotatebox{90}{4. Inc. corrs} & \rotatebox{90}{5. Dec. slopes} & \rotatebox{90}{6. Inc. corrs, dec. slopes}\\
\midrule
 & b = 0 & 94.9 & 94.0 & 94.9 & 95.4 & 94.6 & 94.0 & 95.3 & 93.8 & 95.6 & 95.2 & 95.4 & 95.1 & 94.3 & 95.0 & 94.8 & 95.3 & 94.8 & 94.0\\

 & b = 0.5 & 95.4 & 95.7 & 95.0 & 94.7 & 95.8 & 95.2 & 95.5 & 94.8 & 95.3 & 94.4 & 95.2 & 94.4 & 94.6 & 95.0 & 93.9 & 95.7 & 96.0 & 95.4\\

 & b = 1 & 94.8 & 94.8 & 95.8 & 95.8 & 95.5 & 95.4 & 94.8 & 95.1 & 95.9 & 94.7 & 93.9 & 95.3 & 94.4 & 94.8 & 94.7 & 96.4 & 95.3 & 93.6\\

 & b = 1.5 & 95.4 & 95.1 & 94.2 & 94.9 & 95.4 & 94.9 & 95.2 & 94.9 & 94.2 & 95.4 & 94.1 & 95.2 & 95.4 & 94.8 & 95.2 & 95.1 & 95.4 & 94.8\\

\multirow{-5}{*}{\raggedright\arraybackslash N-HT} & b = 2 & 96.0 & 95.3 & 96.3 & 94.2 & 95.2 & 94.9 & 95.4 & 96.3 & 95.3 & 94.2 & 95.0 & 94.6 & 94.9 & 94.6 & 96.0 & 95.0 & 95.4 & 93.8\\
\cmidrule{1-20}
 & b = 0 & 94.6 & 95.9 & 96.1 & 95.2 & 94.4 & 93.7 & 95.3 & 94.9 & 96.6 & 94.5 & 95.6 & 95.9 & 94.4 & 95.5 & 95.4 & 94.9 & 93.5 & 94.7\\

 & b = 0.5 & 93.3 & 93.6 & 95.6 & 95.3 & 93.5 & 94.6 & 95.1 & 96.1 & 94.0 & 94.3 & 95.3 & 95.0 & 94.6 & 94.6 & 93.8 & 95.4 & 93.7 & 95.6\\

 & b = 1 & 94.6 & 94.9 & 94.5 & 95.0 & 93.5 & 94.6 & 95.2 & 93.5 & 95.9 & 94.7 & 94.9 & 93.0 & 95.7 & 94.7 & 95.0 & 95.1 & 95.1 & 94.4\\

 & b = 1.5 & 95.2 & 93.9 & 94.8 & 95.8 & 96.3 & 94.8 & 93.8 & 94.1 & 94.3 & 94.4 & 95.5 & 93.7 & 95.6 & 96.2 & 95.1 & 95.4 & 94.9 & 95.1\\

\multirow{-5}{*}{\raggedright\arraybackslash C-SR} & b = 2 & 95.3 & 96.2 & 94.5 & 95.4 & 95.3 & 95.2 & 95.2 & 95.9 & 94.7 & 95.0 & 94.8 & 95.2 & 94.4 & 93.6 & 94.8 & 95.1 & 94.7 & 94.9\\
\cmidrule{1-20}
 & b = 0 & 96.1 & 94.5 & 96.4 & 96.1 & 95.4 & 95.4 & 93.8 & 95.4 & 95.3 & 95.8 & 95.7 & 94.0 & 94.5 & 95.7 & 95.2 & 94.5 & 94.8 & 95.7\\

 & b = 0.5 & 94.8 & 95.1 & 95.4 & 95.1 & 95.7 & 95.0 & 95.3 & 95.4 & 96.0 & 94.7 & 95.2 & 94.5 & 95.2 & 96.5 & 94.3 & 95.5 & 94.8 & 94.8\\

 & b = 1 & 95.2 & 95.7 & 95.1 & 95.6 & 95.3 & 95.4 & 95.8 & 95.2 & 96.1 & 95.0 & 96.3 & 96.0 & 95.6 & 96.0 & 95.0 & 95.1 & 94.5 & 96.3\\

 & b = 1.5 & 95.9 & 94.8 & 94.6 & 95.2 & 95.2 & 93.8 & 95.3 & 94.2 & 94.5 & 95.5 & 94.8 & 95.2 & 95.9 & 95.6 & 96.1 & 95.0 & 94.7 & 95.0\\

\multirow{-5}{*}{\raggedright\arraybackslash B-P} & b = 2 & 95.3 & 95.1 & 95.0 & 94.1 & 95.1 & 94.2 & 94.5 & 94.8 & 93.5 & 94.4 & 94.7 & 95.0 & 95.9 & 95.3 & 96.1 & 95.3 & 94.9 & 95.9\\
\bottomrule
\end{tabular}}
\end{table}

		\newpage 
		\begin{table}

\caption{\label{tab:cochran_RT_relbias_coverage}RT-SR results: 1,000*RelBias and CI coverage (\%), by population}
\resizebox{\linewidth}{!}{
\begin{tabular}[t]{l>{}l|rrrrr>{}r|rrrrr>{}r|rrrrr>{}r|}
\toprule
\multicolumn{1}{c}{ } & \multicolumn{1}{c}{ } & \multicolumn{6}{c}{MOS1 (most skewness)} & \multicolumn{6}{c}{MOS2 (mid skewness)} & \multicolumn{6}{c}{MOS3 (least skewness)} \\
\cmidrule(l{3pt}r{3pt}){3-8} \cmidrule(l{3pt}r{3pt}){9-14} \cmidrule(l{3pt}r{3pt}){15-20}
\rotatebox{90}{ } & \rotatebox{90}{ } & \rotatebox{90}{1. Baseline} & \rotatebox{90}{2. Lower corrs} & \rotatebox{90}{3. Higher corrs} & \rotatebox{90}{4. Inc. corrs} & \rotatebox{90}{5. Dec. slopes} & \rotatebox{90}{6. Inc. corrs, dec. slopes} & \rotatebox{90}{1. Baseline} & \rotatebox{90}{2. Lower corrs} & \rotatebox{90}{3. Higher corrs} & \rotatebox{90}{4. Inc. corrs} & \rotatebox{90}{5. Dec. slopes} & \rotatebox{90}{6. Inc. corrs, dec. slopes} & \rotatebox{90}{1. Baseline} & \rotatebox{90}{2. Lower corrs} & \rotatebox{90}{3. Higher corrs} & \rotatebox{90}{4. Inc. corrs} & \rotatebox{90}{5. Dec. slopes} & \rotatebox{90}{6. Inc. corrs, dec. slopes}\\
\midrule
 & b = 0 & -0.6 & -1.1 & 0.1 & -0.0 & -0.4 & -0.6 & -0.0 & -0.1 & 0.0 & -0.0 & -0.0 & 0.4 & 0.2 & 0.1 & -0.1 & 0.4 & 0.1 & -0.2\\

 & b = 1 & 0.0 & 0.2 & 0.2 & -0.2 & -0.6 & 0.1 & 0.3 & 0.8 & -0.0 & -0.1 & -0.0 & -0.4 & 0.6 & 0.1 & -0.0 & 0.4 & -0.2 & -0.1\\

\multirow{-3}{*}{\raggedright\arraybackslash RelBias*1000} & b = 2 & -0.4 & -0.1 & 0.2 & -0.7 & -0.9 & 1.7 & 0.0 & 0.3 & 0.0 & 0.5 & -0.2 & 1.2 & -0.1 & 0.2 & 0.0 & 0.4 & -0.0 & 0.0\\
\cmidrule{1-20}
 & b = 0 & 93.8 & 95.4 & 96.1 & 95.2 & 94.0 & 95.3 & 95.1 & 95.4 & 95.0 & 95.3 & 95.3 & 95.0 & 95.3 & 94.1 & 95.9 & 94.6 & 94.8 & 95.9\\

 & b = 1 & 94.2 & 94.1 & 93.7 & 94.8 & 94.3 & 93.9 & 94.5 & 93.7 & 95.5 & 95.2 & 94.7 & 94.2 & 95.3 & 95.0 & 93.9 & 95.0 & 94.1 & 93.9\\

\multirow{-3}{*}{\raggedright\arraybackslash Coverage (\%)} & b = 2 & 89.8 & 92.2 & 91.3 & 91.6 & 89.2 & 90.7 & 95.7 & 93.9 & 94.3 & 93.3 & 94.5 & 93.9 & 93.8 & 95.3 & 94.2 & 93.3 & 94.1 & 94.7\\
\bottomrule
\end{tabular}}
\end{table}

		\newpage 
		\begin{table}

\caption{\label{tab:Mend_misspec_relbias_coverage}B-P results under model misspecification: 1,000*RelBias and CI coverage (\%) by $b_0$ and $b^{(p)}$ among selected MOS1 populations}
\resizebox{\linewidth}{!}{
\begin{tabular}[t]{l>{}l|>{}r>{}r>{}r>{}r>{}r>{}r|>{}r>{}r>{}r>{}r>{}r>{}r|>{}r>{}r>{}r>{}r>{}r>{}r|}
\toprule
\multicolumn{1}{c}{ } & \multicolumn{1}{c}{ } & \multicolumn{6}{c}{$b^{(p)} = 0$} & \multicolumn{6}{c}{$b^{(p)} = 1$} & \multicolumn{6}{c}{$b^{(p)} = 2$} \\
\cmidrule(l{3pt}r{3pt}){3-8} \cmidrule(l{3pt}r{3pt}){9-14} \cmidrule(l{3pt}r{3pt}){15-20}
\rotatebox{90}{ } & \rotatebox{90}{ } & \rotatebox{90}{1. Baseline} & \rotatebox{90}{2. Lower corrs} & \rotatebox{90}{3. Higher corrs} & \rotatebox{90}{4. Inc. corrs} & \rotatebox{90}{5. Dec. slopes} & \rotatebox{90}{6. Inc. corrs, dec. slopes} & \rotatebox{90}{1. Baseline} & \rotatebox{90}{2. Lower corrs} & \rotatebox{90}{3. Higher corrs} & \rotatebox{90}{4. Inc. corrs} & \rotatebox{90}{5. Dec. slopes} & \rotatebox{90}{6. Inc. corrs, dec. slopes} & \rotatebox{90}{1. Baseline} & \rotatebox{90}{2. Lower corrs} & \rotatebox{90}{3. Higher corrs} & \rotatebox{90}{4. Inc. corrs} & \rotatebox{90}{5. Dec. slopes} & \rotatebox{90}{6. Inc. corrs, dec. slopes}\\
\midrule
 & $b_0 = 0$ & \cellcolor{lightgray}{1.0} & \cellcolor{lightgray}{1.1} & \cellcolor{lightgray}{0.4} & \cellcolor{lightgray}{-1.5} & \cellcolor{lightgray}{0.2} & \cellcolor{lightgray}{2.4} & \cellcolor{white}{-0.0} & \cellcolor{white}{0.6} & \cellcolor{white}{0.1} & \cellcolor{white}{0.2} & \cellcolor{white}{0.0} & \cellcolor{white}{-0.2} & \cellcolor{white}{-0.7} & \cellcolor{white}{0.6} & \cellcolor{white}{0.5} & \cellcolor{white}{2.7} & \cellcolor{white}{1.9} & \cellcolor{white}{2.3}\\

 & $b_0 = 0.5$ & \cellcolor{white}{0.9} & \cellcolor{white}{-0.6} & \cellcolor{white}{0.5} & \cellcolor{white}{-1.6} & \cellcolor{white}{0.5} & \cellcolor{white}{0.8} & \cellcolor{white}{-0.6} & \cellcolor{white}{0.1} & \cellcolor{white}{0.2} & \cellcolor{white}{0.4} & \cellcolor{white}{0.9} & \cellcolor{white}{-0.8} & \cellcolor{white}{-0.5} & \cellcolor{white}{2.3} & \cellcolor{white}{0.7} & \cellcolor{white}{3.1} & \cellcolor{white}{1.7} & \cellcolor{white}{2.0}\\

 & $b_0 = 1$ & \cellcolor{white}{1.0} & \cellcolor{white}{-0.7} & \cellcolor{white}{0.3} & \cellcolor{white}{-1.3} & \cellcolor{white}{-0.3} & \cellcolor{white}{2.0} & \cellcolor{lightgray}{-0.2} & \cellcolor{lightgray}{-0.4} & \cellcolor{lightgray}{-0.1} & \cellcolor{lightgray}{0.0} & \cellcolor{lightgray}{0.0} & \cellcolor{lightgray}{1.1} & \cellcolor{white}{-0.4} & \cellcolor{white}{1.3} & \cellcolor{white}{0.4} & \cellcolor{white}{2.9} & \cellcolor{white}{2.5} & \cellcolor{white}{2.6}\\

 & $b_0 = 1.5$ & \cellcolor{white}{2.2} & \cellcolor{white}{-1.8} & \cellcolor{white}{0.1} & \cellcolor{white}{-3.4} & \cellcolor{white}{0.4} & \cellcolor{white}{-0.1} & \cellcolor{white}{0.0} & \cellcolor{white}{0.2} & \cellcolor{white}{-0.1} & \cellcolor{white}{-0.1} & \cellcolor{white}{-0.0} & \cellcolor{white}{-0.2} & \cellcolor{white}{-0.6} & \cellcolor{white}{1.5} & \cellcolor{white}{0.4} & \cellcolor{white}{3.0} & \cellcolor{white}{1.5} & \cellcolor{white}{2.3}\\

\multirow{-5}{*}{\raggedright\arraybackslash RelBias*1000} & $b_0 = 2$ & \cellcolor{white}{1.6} & \cellcolor{white}{-3.8} & \cellcolor{white}{-0.5} & \cellcolor{white}{-1.8} & \cellcolor{white}{0.8} & \cellcolor{white}{3.1} & \cellcolor{white}{-0.2} & \cellcolor{white}{0.2} & \cellcolor{white}{0.1} & \cellcolor{white}{-0.4} & \cellcolor{white}{-0.3} & \cellcolor{white}{-0.1} & \cellcolor{lightgray}{-0.4} & \cellcolor{lightgray}{2.3} & \cellcolor{lightgray}{0.6} & \cellcolor{lightgray}{2.9} & \cellcolor{lightgray}{1.6} & \cellcolor{lightgray}{2.4}\\
\cmidrule{1-20}
 & $b_0 = 0$ & \cellcolor{lightgray}{96.1} & \cellcolor{lightgray}{94.5} & \cellcolor{lightgray}{96.4} & \cellcolor{lightgray}{96.1} & \cellcolor{lightgray}{95.4} & \cellcolor{lightgray}{95.4} & \cellcolor{white}{95.8} & \cellcolor{white}{94.6} & \cellcolor{white}{95.0} & \cellcolor{white}{93.4} & \cellcolor{white}{95.5} & \cellcolor{white}{93.8} & \cellcolor{white}{96.0} & \cellcolor{white}{95.5} & \cellcolor{white}{95.8} & \cellcolor{white}{94.9} & \cellcolor{white}{93.7} & \cellcolor{white}{95.6}\\

 & $b_0 = 0.5$ & \cellcolor{white}{94.3} & \cellcolor{white}{94.8} & \cellcolor{white}{95.1} & \cellcolor{white}{95.0} & \cellcolor{white}{96.1} & \cellcolor{white}{94.1} & \cellcolor{white}{94.3} & \cellcolor{white}{95.1} & \cellcolor{white}{95.3} & \cellcolor{white}{94.4} & \cellcolor{white}{94.7} & \cellcolor{white}{95.1} & \cellcolor{white}{94.9} & \cellcolor{white}{95.2} & \cellcolor{white}{95.1} & \cellcolor{white}{94.6} & \cellcolor{white}{95.8} & \cellcolor{white}{96.6}\\

 & $b_0 = 1$ & \cellcolor{white}{93.7} & \cellcolor{white}{95.3} & \cellcolor{white}{94.8} & \cellcolor{white}{95.9} & \cellcolor{white}{95.4} & \cellcolor{white}{95.8} & \cellcolor{lightgray}{95.2} & \cellcolor{lightgray}{95.7} & \cellcolor{lightgray}{95.1} & \cellcolor{lightgray}{95.6} & \cellcolor{lightgray}{95.3} & \cellcolor{lightgray}{95.4} & \cellcolor{white}{95.0} & \cellcolor{white}{95.2} & \cellcolor{white}{94.9} & \cellcolor{white}{94.6} & \cellcolor{white}{95.2} & \cellcolor{white}{94.8}\\

 & $b_0 = 1.5$ & \cellcolor{white}{97.0} & \cellcolor{white}{96.5} & \cellcolor{white}{96.7} & \cellcolor{white}{97.1} & \cellcolor{white}{95.7} & \cellcolor{white}{98.1} & \cellcolor{white}{95.7} & \cellcolor{white}{95.5} & \cellcolor{white}{95.5} & \cellcolor{white}{94.7} & \cellcolor{white}{95.6} & \cellcolor{white}{95.1} & \cellcolor{white}{94.0} & \cellcolor{white}{96.2} & \cellcolor{white}{96.4} & \cellcolor{white}{94.6} & \cellcolor{white}{96.3} & \cellcolor{white}{95.0}\\

\multirow{-5}{*}{\raggedright\arraybackslash Coverage (\%)} & $b_0 = 2$ & \cellcolor{white}{98.6} & \cellcolor{white}{98.1} & \cellcolor{white}{98.3} & \cellcolor{white}{98.2} & \cellcolor{white}{98.8} & \cellcolor{white}{98.1} & \cellcolor{white}{96.1} & \cellcolor{white}{96.0} & \cellcolor{white}{95.9} & \cellcolor{white}{95.6} & \cellcolor{white}{95.4} & \cellcolor{white}{97.6} & \cellcolor{lightgray}{95.3} & \cellcolor{lightgray}{95.1} & \cellcolor{lightgray}{95.0} & \cellcolor{lightgray}{94.1} & \cellcolor{lightgray}{95.1} & \cellcolor{lightgray}{94.2}\\
\bottomrule
\multicolumn{20}{l}{\rule{0pt}{1em}\textit{Note: } Gray background denotes use of the correct specification (i.e., $b_0 = b^{(p)}$).}\\
\end{tabular}}
\end{table}

	\end{singlespace}
\end{landscape}

\begin{figure}[H]
	\centering
	\caption{Q-Q plots for unstratified regressions of IRS data, based on log-transformed and original scales}
	\label{fig:app_irs_qqplot_faceted} 
	\includegraphics[scale=1]{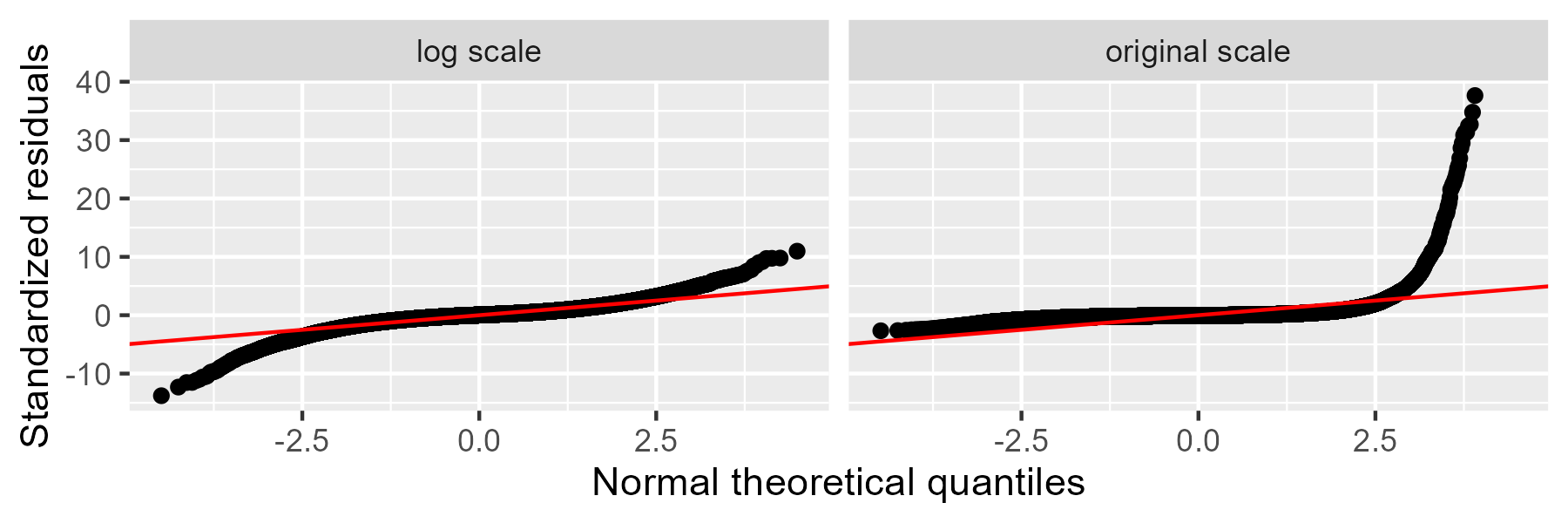}
	\captionsetup{font=it,singlelinecheck=off}
	\caption*{Note. Lefthand panel is a Q-Q plot for an unstratified, no-intercept simple linear regression of the NCCS data from Section 6.1 of the main paper, %\S\ref{sec:ch2:NCCS_simulation},  %NOTE: refers specifically to the log-scale sim
	where $X_i$ and $Y_i$ denote log revenues of unit $i$ for 2008 and 2013, respectively, with analytical weights of by $1/X_i^{0.55}$. %, for $i = 1, 2, ..., 140,793$.
		Righthand panel is an analogous Q-Q plot for the NCCS data on the original scale, as in Section 6.2 of the main paper, %\S\ref{sec:ch2:NCCS_simulation_origscale}, 
		meaning that $X_i$ and $Y_i$ denote revenues of unit $i$ for 2008 and 2013, respectively, and with analytical weights of $1/X_i^{1.19}$. %, for $i = 1, 2, ..., 138,960$.
		Righthand panel does not display six observations with standardized residuals greater than 40.}
\end{figure}

\begin{singlespace}
	
\begin{longtable}[t]{>{}l|rr>{}r|r}
\caption{NCCS data: population counts by sector and 2008 revenue (with size classes based on original scale)}\\
\toprule
\multicolumn{1}{c}{ } & \multicolumn{3}{c}{Total Revenue in 2008} & \multicolumn{1}{c}{ } \\
\cmidrule(l{3pt}r{3pt}){2-4}
Nonprofit Sector & Under \$1.8m & \$1.8m--\$12m & \$12m--\$100m & Total\\
\midrule
\endfirsthead
\caption[]{NCCS data: population counts by sector and 2008 revenue (with size classes based on original scale) \textit{(continued)}}\\
\toprule
\multicolumn{1}{c}{ } & \multicolumn{3}{c}{Total Revenue in 2008} & \multicolumn{1}{c}{ } \\
\cmidrule(l{3pt}r{3pt}){2-4}
Nonprofit Sector & Under \$1.8m & \$1.8m--\$12m & \$12m--\$100m & Total\\
\midrule
\endhead

\endfoot
\bottomrule
\endlastfoot
Arts, culture, and humanities & 12,239 & 1,709 & 314 & 14,262\\
Education & 12,996 & 4,507 & 1,646 & 19,149\\
Environment & 5,246 & 781 & 118 & 6,145\\
Health & 11,675 & 5,755 & 3,185 & 20,615\\
Human services & 44,088 & 10,824 & 2,609 & 57,521\\
International & 2,332 & 482 & 176 & 2,990\\
Public and societal benefit & 8,740 & 1,702 & 366 & 10,808\\
Religion & 6,681 & 701 & 88 & 7,470\\
\midrule
Total & 103,997 & 26,461 & 8,502 & 138,960\\*
\end{longtable}

	{\makeatletter\def\@currentlabel{\thetable}\label{table:irs_popsizes_sim_100m}}
\end{singlespace}

\end{document}

\typeout{get arXiv to do 4 passes: Label(s) may have changed. Rerun}